\newif\iffigs
  \figsfalse
\documentclass[12pt]{article}
\usepackage{latexsym,amssymb}
\iffigs
    \usepackage{graphicx}
    \input{epsf}
\else
\fi
\newcommand{\ft}[2]{{\textstyle\frac{#1}{#2}}}

\newsavebox{\uuunit}
\sbox{\uuunit}
                             {\setlength{\unitlength}{0.825em}
                                      \begin{picture}(0.6,0.7)
                                                                  \thinlines
                                                                  \put(0,0){\line(1,0){0.5}}
                                                                  \put(0.15,0){\line(0,1){0.7}}
                                                                  \put(0.35,0){\line(0,1){0.8}}
                                                                 \multiput(0.3,0.8)(-0.04,-0.02){10}{\rule{0.5pt}{0.5pt}}
                                      \end {picture}}

\makeatletter \@addtoreset{equation}{section} \makeatother

\newtheorem{definizione}{Principle}[section]
\def\bfone{\relax{\rm 1\kern-.35em 1}}

\begin{document}
\begin{titlepage}
\begin{flushright}
CERN-PH-TH/2006-113\\
DFTT/10-2006\\
DISTA-2006 \\
hep-th/0606171
\end{flushright}
\vskip 1.5cm
\begin{center}
{\LARGE \bf Pure Spinors, Free Differential Algebras,\\
\vskip 0.2cm
and the Supermembrane$^\dagger$
%
} \\
 \vfill
{\large Pietro Fr{\'e}$^1$ and  Pietro Antonio Grassi$^2$} \\
\vfill {
$^1$ Dipartimento di Fisica Teorica, Universit{\`a} di Torino, \\
$\&$ INFN -
Sezione di Torino\\
via P. Giuria 1, I-10125 Torino, Italy\\
\vskip .3cm
$^2$ { Centro Studi e Ricerche E. Fermi,
Compendio Viminale, I-00184, Roma, Italy,}\\
{ DISTA, Universit\`a del Piemonte Orientale, }\\
{ Via Bellini 25/G,  Alessandria, 15100, Italy 
$\&$ INFN - Sezione di
Torino}\\
{CERN, Theory Unit, CH-1211 Geneve, 23, Switzerland}}
\end{center}
\vfill
\begin{abstract}
The lagrangian formalism for the supermembrane in any 11d supergravity
background is constructed in the pure spinor framework. Our gauge-fixed action
is manifestly BRST, supersymmetric, and 3d Lorentz invariant. The relation between
the Free Differential Algebras (FDA) underlying  11d supergravity and the BRST
symmetry of the membrane action is exploited. The "gauge-fixing" has a natural
interpretation as the variation of the Chevalley cohomology class needed for
the extension of 11d super-Poincar\'e superalgebra to M-theory FDA. We study
the solution of the pure spinor constraints in full detail.
\end{abstract}
\vfill
\vspace{1.5cm}
\vspace{2mm} \vfill \hrule width 3.cm {\footnotesize $^ \dagger $
This work is supported in part by the European Union RTN contract
MRTN-CT-2004-005104 and by the Italian Ministry of University (MIUR) under
contracts PRIN 2005-024045 and PRIN 2005-023102}
\end{titlepage}


\section{Introduction}
\label{introibo}
The strong regime of string theory is usually denoted M-theory. However,
up to now, the underlying fundamental theory and the degrees of freedom are
still unknown. Some indications coming from the low-energy effective action,
accurately described by 11-dimensional supergravity, from the presence of extended objects in string theory
such
as the supermembrane and the M5-brane (from which all D-branes can be obtained by dimensional
reduction), and from the superalgebra in 11 dimensions, pointed out that a plausible
candidate for the fundamental theory is the theory of the {\it supermembrane}.
This theory has been discovered in \cite{BST} and, since then,
several studies have been performed to understand if it
really has all the necessary features to describe M-theory. We do not present
here a review and we refer to \cite{review} for a complete account on the subject.
Anyway, we need to remind the reader of some basic facts about the supermembrane.
\par
The supermembrane is a theory of maps from a (2+1)-worldvolume to a
(10+1)-dimen\-sional target space. When the membrane moves in a flat superspace,
the fundamental mathematical quantities are the supersymmetric line elements $\Pi^{\underline a}_{i} =
\, e^\mu_i(\xi) \, \left (\partial_{\mu} X^{\underline a} +  {\rm i} \,\bar\theta \Gamma^{\underline a}
\partial_{\mu} \theta \right)$ and $\psi_i ^{\underline{\alpha}} \, = \, e^\mu_i(\xi) \left( \partial_{\mu} \theta^{\underline \alpha}\, \right ) $
where $i=1,2,3$ are the worldvolume flat indices, ($e^\mu_i (\xi)$ is the inverse dreibein) $\underline a=0, \dots,
10$ are the flat target space indices and $\alpha=1, \dots, 32$ are
the indices for the spinorial representation of $\mathrm{SO(1,10)}$. The coordinates $X^{\underline a}$ and
$\theta^{\underline \alpha}$ define a local basis in the superspace of the target space. The theory
written in terms of the basic supersymmetric quantities is manifestly Lorentz and supersymmetric
invariant. In addition, it is invariant under local diffeomorphism on the worldvolume and
under an infinite reducible gauge symmetry known as $\kappa$-symmetry (see \cite{BST,ksusyschwarz}
and references therein for details). These gauge symmetries remove the correct number of
bosonic and fermionic degrees of freedom to have a manifestly supersymmetric spectrum \cite{spectrum} and
they are crucial in the quantization procedure.
\par
The symmetries (rigid and local), the action and the supersymmetry of the supermembrane theory
are very similar to those of the superstring in the Green-Schwarz (GS) formalism and therefore
we can use it as an example. The GS superstring
is characterized by a set of fermionic constraints $d_{\underline \alpha}$
of the first and second class type. These constraints are entangled together in a single quantity
\begin{equation}
  d_{\underline{\alpha}} \equiv p_{\underline{\alpha}} - \frac{\partial {\cal L}_{GS}}
  {\partial_{0} \theta^{\underline{\alpha}}} =0
\label{dA-formula}
\end{equation}
where ${\cal L}_{GS}$ is the Green-Schwarz action
and it is impossible to separate them locally without
breaking Lorentz invariance and supersymmetry.
\par
To follow the conventional quantization strategy, one has to use
the BRST quantization technique for the first class constraints and
to replace the Poisson brackets by the Dirac bracket to
take into account the second class constraints.  Instead,
following Berkovits \cite{Berkovits:2000fe}, one defines a BRST-like
charge by
\begin{equation}
  Q = \int_{\xi_{0} = t} d^{d-1}\xi \, \left( \lambda^{\underline \alpha}
  d_{\underline \alpha} \right)(\xi)\,,
\label{Qberkovo}
\end{equation}
where $\lambda^{\underline\alpha}$ are commuting spinors  and
$\xi^{\mu}$ are the worldvolume coordinates.
The integration is extended over the spatial coordinates.
Here we make no distinction between first and
second class constraints and the Poisson brackets are used to compute the
commutation relations. This implies that for a nilpotent BRST symmetry, some constraints
on the ghost fields $\lambda^{\underline \alpha}(\xi)$ are necessary and
the latter are now known in the literature as {\it pure spinor constraints}.
\par
During the last six years, the quantization of superstring according to
\cite{Berkovits:2000fe} has been studied and several results have been already
achieved. Still, the formalism is not yet complete and several issues need to be understood and
clarified.  One of these issues is the geometric structure underneath: even if the
formalism seems to give consistent results, it is rather important to spell out
its geometrical structure in particular to study $M2$-branes in a generic 11d supergravity background. 
As an example,
the geometrical formulation of superstrings and supermembranes using $\kappa$-symmetry
like in ref. \cite{BST} permits useful expressions for any worldvolume and target space background.
\par
The pure spinor formalism for the superstrings has been adapted to supermembranes
in \cite{Berkovits:2002uc}. There it is shown that starting from the
original action of \cite{BST}, one can derive the fermionic constraints $d_A(\xi)$
from which the BRST charge $Q$ can be constructed.  The ghost fields $\lambda^A(\xi)$
carry an index in $\mathrm{Spin(32)}
$, they are commuting scalars and
they satisfy to pure spinor conditions
\begin{equation}
 \bar\lambda  \Gamma^{\underline a} \lambda =0\,,
\quad\quad
\bar\lambda\Gamma^{\underline a b}\lambda \eta_{\underline a b}\Pi^{\underline b}_I =0\,,
\quad\quad
\bar\lambda \partial_I \lambda =0 \,,
\label{orfana1}
\end{equation}
where the index $I$ only runs over the spatial directions of the worldvolume.
\par
In \cite{Berkovits:2002uc} the supermembrane is studied
using the Hamiltonian approach \cite{Bergshoeff:1987in} and therefore it is not
manifestly covariant on the worldvolume. Nevertheless, it is shown that by freezing the
transverse degrees of freedom of the membrane, the action reduces to a
superparticle action whose spectrum is identified with 11d supergravity (and the
equations of motion are given at the linearized level). Therefore, the pure spinor supermembrane action has
all symmetries gauge-fixed and it provides the starting point for a complete analysis at the quantum level.
\par
Unfortunately, even in this new framework the gauge fixed action is interacting and it cannot
be  further simplified. Therefore, besides the first massless level, the full spectrum
of the membrane is still unknown (see \cite{Duff:1987cs} and \cite{spectrum} for a
review on the spectrum of the supermembrane in the semiclassical regime and on the light-cone gauge).
It is however useful to note that some computations can be indeed performed explicitly.
By analyzing the low-energy spectrum, one can discover that the states are organized in such a way that
some tree level and one-loop amplitudes can be constructed (almost algebraically by zero
mode saturation rules  \cite{Berkovits:2004px}) and computed in the
approximation of zero transverse fluctuations (\cite{Anguelova:2004pg} and \cite{Mafra:2005jh}).
\par
In \cite{Berkovits:2002uc} (following \cite{Oda:2001zm}),
the pure spinor action is obtained by a BRST-like approach where the classical action is replaced by the
gauge fixed action by adding a BRST-exact term
\begin{equation}
  S_{classical} \rightarrow S_{Classical}\, + \, {\cal S} \int d^{d} \xi \,  \Phi_{gauge}(\xi)
\label{orfana2}
\end{equation}
where the \textit{gauge fermion} $\Phi_{gauge}(\xi)$ is a local functional of the fields of the theory and
we denote by ${\cal S}$ the functional differential BRST operator. In the
conventional BRST approach the quantum action is invariant under the BRST symmetry since
the classical action is invariant under the gauge symmetry and the BRST operator is nilpotent.
In the pure spinor string/membrane theory, each single term is not invariant, but only the sum is such.
This is possible because of the non-invariance of the classical action and of the
non-nilpotency of the BRST charge (this point will be elucidated in the text).
\par
The form of the BRST charge is obtained by using the Hamiltonian formalism, and
therefore the action of the charge on each field is determined by the Poisson brackets
of the charge with the corresponding fields.
\par
In the present work, we rather start from a Lagrangian approach. In that case the action of the
BRST charge on the conjugate
momenta is not fixed {\it a priori}, but it should be determined from the consistency of the formalism.
Therefore the starting point is rather different and, as we are going to emphasize leads to
a full fledged geometrical interpretation of the action, of the BRST transformations and also of the
pure spinor constraints.
\par
In order to introduce the reader to  our viewpoint we have to recall some
fundamental developments in the general geometrical understanding of
supersymmetric field theories which are by now quite old, but they are quite essential to our present arguing. The first is the geometrical decoding
of local supersymmetry itself.

\subsection{Geometry of Supersymmetry}

In every supergravity theory for any
number of permitted space--time dimensions ($2 \le D \le 11$) and for
any number of permitted supersymmetry charges $\mathcal{N}_{SUSY}$ a
supersymmetry transformation is nothing else but a \textit{Lie
derivative} $ \mathcal{L}_{\overrightarrow{\epsilon}} $ along a tangent vector ${\overrightarrow{\epsilon}}$
of fermionic type. The crucial and fascinating point is however
that the superfields describing the geometry of superspace
on which such Lie derivatives act -- namely the  supervielbein and the  super-connection appropriate to the considered case --
are not free, rather they have to satisfy a unique set of constraints similar
to Cauchy--Riemann conditions.
These are named the \textit{rheonomic conditions} \cite{castdauriafre} and
encode all the symmetries together with the classical dynamics of any
supergravity. In short, they are the very definition of the theory.
Mathematically the rheonomic conditions
(i.e.\,\textit{rheonomic principle}) are expressed by the requirement that the
components of all superspace curvature components along fermionic directions
should be expressed as linear combinations of the curvature
components along bosonic directions. Obviously this must be done in a
way compatible with the fulfillment of Bianchi identities and it
turns out that the solution to such a problem is always unique. It
determines the explicit form of the supersymmetry transformations on
all fields of the theory and at the same time it also determines the
dynamics. Indeed, the rheonomic conditions imply some constraints on the bosonic curvature
components that are interpreted as the field equations (Einstein
equation, Rarita-Schwinger equations and so on). Therefore, supergravity
theories are completely determined by the choice of the superalgebra
plus the construction of the unique \textit{rheonomic
parametrization} of its curvatures. In the case of higher dimensional
supergravity superalgebras are replaced by the larger category of
Free Differential Algebras (See appendix \ref{appA} for a short review of the concept.)
\par
The second fundamental  advance relevant to our present
discussion is the geometrical interpretation of $\kappa$-symmetry in
$p$-brane theories. Indeed it was realized that this is no new exotic symmetry which has to be
invented case by case rather it is nothing else, but the same
supersymmetry which is already determined by the \textit{unique
rheonomic parametrization} of the supergravity curvatures
describing the ambient superspace geometry in which the $p$--brane
evolves. The only novelty is a restriction on the fermionic tangent
vector $\overrightarrow{\epsilon}$ along which one can calculate the
Lie derivative $L_{\overrightarrow{\epsilon}}$. This restriction is
actually encoded in a projection operator $\mathcal{P}_{q}$, which
applied to $\overrightarrow{\epsilon}$ enforces the operation $L_{\mathcal{P}_{q}\overrightarrow{\epsilon}}$
to be parallel to the world volume evolution. This viewpoint on
$\kappa$-symmetry is very fruitful. It originated from work done in
\cite{ksusytonin} and was extended and fully merged with rheonomy
in \cite{Dall'Agata:1999wz,Fre:2002kc}. In this latter paper, in particular, it was observed that the explicit
form of $\kappa$-symmetry, namely the projector $\mathcal{P}_{q}$  can
be  easily derived from a first-order action coupled to a generic supergravity background
and the general rules were established how to write
$\kappa$-symmetric $p$-brane actions in first order rheonomic
formalism. The case of the $\mathrm{M2}$--brane was spelled out explicitly \cite{Dall'Agata:1999wz}
and
it was shown to lead to the anti de Sitter supersingleton in case the
background is chosen to be $\mathrm{AdS}_4 \times \mathbb{S}^7$.
\par
The third advance in geometrical understanding concerns the geometrical decoding of the BRST quantization of supergravity theories. The first step due to \cite{Baulieu:1986dp}
consists of constructing a double elliptic complex in which the
exterior derivative operator $d$ in superspace and the $BRST$ charge
$Q=\mathcal{S}$ are merged into a new nilpotent operator $\mathbf{d}
= d + \mathcal{S}$, while all the super $p$-forms are extended to super
$p$ ghost-forms, the extra components of which are the ghost-fields.
In order to obtain explicit $\mathrm{BRST}$-transformations, however, one has
to implement suitable conditions on the curvatures of the ghost-forms, similar to the
rheonomic conditions imposed on classical super curvatures in order to
define supergravity. In \cite{Anselmi:1992tj} it was discovered that
the right answer encompassing the correct BRST algebra for any
supergravity theory is provided by a very simple and general
principle which states the following \textit{The rheonomic parametrization of the ghost-form
supercurvatures is formally identical, mutatis mutandis,  to the  rheonomic
parametrization of the classical supercurvatures}. In other words it
suffices to keep the same form for all  the curvature components
and simply substitute the extended ghost-form where the corresponding
classical form appeared.

\subsection{Outlook}

In the present paper, relying on the above results as starting point we
take a leap forward and we show where the pure spinor constraints come
from. They  emerge from a constrained BRST algebra which is
obtained from the ordinary BRST algebra of $D=11$ supergravity
by imposing that the diffeomorphism ghosts and the gauge ghosts of the three-form $\mathbf{A}^{[3]}$ should be zero, namely
by unfreezing the target space diffeomorphisms and the three-form gauge transformations. The logical steps
are the following ones:
\begin{enumerate}
  \item The superPoincar\'e algebra in $D=11$, uniquely defines its own extension to a Free
  Differential Algebra $\mathrm{FDA}_{11}$ via its cohomology (see appendix
  \ref{appA}).
  \item The $\mathrm{FDA}_{11}$ via its unique rheonomic parametrization defines classical D=11
  supergravity and, in conjunction with the classical supermembrane action, defines the $\kappa$-symmetry
  transformations of the latter.
  \item The rheonomic parametrization of the $\mathrm{FDA}_{11}$
  curvatures, uniquely extended to ghost-forms define the
  unconstrained $\mathrm{BRST}$-algebra of $D=11$ supergravity. Under these
  $\mathrm{BRST}$-transformation the $\kappa$-symmetric classical
  supermembrane action is not invariant.
  \item The constraint that the diffeomorphisms ghosts should be
  zero, inserted in the ordinary $\mathrm{BRST}$ algebra uniquely determines, from consistency, a
  set of constraints on the superghosts that are identified as the
  \textit{primary pure spinor constraints}. The supermembrane action
  can be made invariant against this constrained $\mathrm{BRST}$ algebra by
  adding to the classical action a new uniquely determined
  \textit{gauge fixing part} of the form $\mathcal{S} \int d^{d} \xi \,
  \Phi_{gauge}(\xi)$ which is actually related to the very same
  cohomology class which originated the FDA extension and all the
  rest.
\end{enumerate}
So, we start from the target space BRST-algebra and we set to zero both the translation and the gauge symmetry ghosts
of the three-form and we require that the algebra still closes. Explicitly this implies that the supersymmetry
ghost $\lambda$ should satisfy the following constraints:
\begin{equation}
\bar\lambda \Gamma^{\underline a} \lambda =0\,, \quad\quad
\bar\lambda \Gamma^{\underline a b} \lambda
\eta_{\underline a a'} \eta_{\underline b b'}
\Pi^{\underline a'}_{[i} \Pi^{\underline b'}_{j]}=0\,.
\label{orfanella4}
\end{equation}
It is easy to compare these constraints with those found using the Hamiltonian formalism: these are
weaker, but the first constraint is just the usual 11d pure spinor constraint for the superparticle.
\par
>From supersymmetry, we can easily deduce the BRST transformation rules for the
fields $X^{\underline a}, \theta^{\underline \alpha}, \dots$ and we can compute the BRST variation of the action. This
variation turns out to be proportional to the gravitino form $\Psi$ and as we already stated
it should be cancelled by a suitable
gauge fixing term. The gauge fixing term is guessed of the form $\mathcal{S}\left(something\right)$ and, therefore,
its variation comes only from the nilpotency of the charge itself. We show that it is indeed possible to
find a suitable gauge fixing and a BRST variation of the antighost field to have an invariant
action.
\par
As a second step, we have to check the nilpotency and the consistency condition
for the BRST transformation rules. We found that under a stronger form of the constraints
derived from the supersymmetry algebra the BRST algebra closes and we have a consistent
framework. It is important to note that the constraint we found are completely covariant
on the worldvolume and a complete solution is given in order to
show that they are not empty (in the previous publications \cite{Berkovits:2002uc} and
\cite{Aisaka:2006by} only a partial analysis was performed). Finally, we show
that, quite remarkably, as we already anticipated, this term is unique and related
to the basic cohomological class of the $D=11$ superPoincar\'e
algebra. The application of geometrical techniques to the quantization of superstrings 
were already performed in series of works \cite{GrassiandPeter} culminating with a 
reformulation of the pure spinor formalism in terms of WZW model where the coset 
is gauged by means of the pure spinor BRST charge. 
A similar analysis was done for the D-branes in \cite{Anguelova:2003sn}.
\par
The paper is organized as follows: in section \ref{classM2}, we review some basic facts about the classical
action, the $\kappa$-symmetry, the first-order formalism and the so named double first order formalism
\cite{Fre:2002kc}, which is obligatory in order to include world-volume gauge fields so as to obtain $\kappa$-symmetric
actions of the Born-Infeld type and is based on the extension of the $\mathrm{SO(1,d)}$ Lorentz symmetry to a local
$\mathrm{GL(d,\mathbb{R})}$ symmetry. This point will be relevant in the
extension of our pure spinor geometrical quantization from the case of the $M2$--brane to the case of
$Dp$-branes.
In section \ref{BRSgen}, we discuss the BRST quantization of 11d supergravity, the rheonomic parametrization of
its curvatures, and the ensuing BRST symmetry with the pure spinor constraints.
In section \ref{gaugiofisso}, the gauge fixing
and the complete BRST is constructed. The nilpotency is discussed and the
pure spinor constraints are solved in a well-adapted basis. In section \ref{exempla}, the membrane action
constructed on a generic background is exemplified on two instances of specific backgrounds.
Section \ref{concludo} is devoted to the conclusions and to new developments.
In Appendix \ref{appA} we recall some definitions of the Free Differential Algebras.


\section{The classical supermembrane 
}
\label{classM2}
In the context of superstrings and  of other
$p$--brane classical theories the important issue is to write world--volume
actions that possess both reparametrization invariance and $\kappa$--symmetry
\cite{ksusyschwarz}. The former is needed to
remove the unphysical degrees of freedom of the bosonic sector, while the
latter removes the unphysical fermions. In this way we end up with an equal number of
physical bosons and physical fermions as it is required by supersymmetry.
As widely discussed in the literature \cite{ksusytonin,castdauriafre,Dall'Agata:1999wz,allroads}
the appropriate $\kappa$--symmetry transformation rules are
nothing else but the supersymmetry transformation rules of the bulk
supergravity background fields with a special supersymmetry parameter
$\epsilon$ that is projected onto the brane. For those $\kappa$-supersymmetric branes where
the gauge field strength $F_{\mu \nu }$ is not required (for example the string itself or the supermembrane)
such a projection is realized by imposing that the spinor $\epsilon$
satisfies the following condition:
\begin{equation}
  \epsilon = \ft {1}{2} \left( 1 + \left( {\rm i}\right) ^{d}\, \ft{1}{d!} \, \Gamma_{\underline{a_1\dots a_d}}
  V^{\underline{a_1}}_{i_1} \, \dots \,
  V^{\underline{a_d}}_{i_d}\,\epsilon^{i_1\dots i_d} \right)
  \epsilon
\label{projparam}
\end{equation}
where $\Gamma_{\underline{a}}$ are the gamma matrices in
$D$--dimensions and $V^{\underline{a}}_{m}$ are the component of
the bulk vielbein $V^{\underline{a}}$ onto a basis of world-volume
vielbein $e^m$. Explicitly we write
\begin{equation}
  V^{\underline{b}}_m \, e^m = \varphi^* \, \left[ V^{\underline{b}} \right]
\label{pullobacco}
\end{equation}
where $\varphi^* \, \left[ V^{\underline{b}} \right]$ denotes  the pull--back of the bulk vielbein on the worldvolume,
\begin{equation}
  \varphi \, : \, W_d \, \hookrightarrow \,  \mathcal{M}_D
\label{inietto}
\end{equation}
being the injection map of the worldvolume  $W_d$
into the target space $\mathcal{M}_D$.
\par
It was shown in \cite{castdauriafre} and explicitly applied to the
 case of the supermembrane in
\cite{Dall'Agata:1999wz} that by using a first order
formalism on the worldvolume  the implementation of
$\kappa$--symmetry is reduced to an almost trivial matter once the rheonomic
parametrizations, consistent with superspace
Bianchi identities,  are given for all the the curvatures of the bulk background fields.
\subsection{The first order form of the kinetic term for a
$p$-brane}
\label{oldfirst}
The first order formulation of the Nambu--Goto action\cite{nambugotorig}  is the Polyakov action
\cite{polyakov} for $p$--branes:
\begin{equation}
  \mathcal{L}^{Polyakov}_{p-brane} = \frac {1}{2(d-1)} \, \int \, d^d\xi
  \, \sqrt{ - \, \mbox{det}\, h_{\mu \nu }} \, \left\{ \, h^{\rho \sigma
  } \, \partial _\rho  X^{\underline{\mu} } \, \partial _\sigma  X^{\underline{\nu}
  } \, g_{\underline{\mu \nu} } + \left( d-2 \right) \right\}
\label{Polyak}
\end{equation}
where the auxiliary field $h_{\rho \sigma }$ denotes the world--volume
metric. Varying the action (\ref{Polyak}) with respect to $\delta h_{\rho \sigma
}$ we obtain the equation:
\begin{equation}
  h_{\rho \sigma } = G_{\rho \sigma } \equiv \partial _\rho X^{\underline{\mu}} \, \partial _\sigma
  X^{\underline{\nu}}\, g_{\underline{\mu \nu} }
\label{identh}
\end{equation}
and substituting (\ref{identh}) back into (\ref{Polyak}) we retrieve
the second order Nambu Goto action.
\par
The Polyakov action (\ref{Polyak}) is not yet in a suitable form for
a simple geometric implementation of $\kappa$--symmetry, but can
be easily converted to such a form. The required steps are:
\begin{enumerate}
  \item replacing the world--volume metric $h_{\mu \nu }(\xi)$ with a
  world--volume vielbein $e^i=e^i_\rho \, d\xi^\rho$,
  \item using a first order formalism also for the derivatives of
  target space coordinates $X^{\underline{\mu}}$ with respect to the
  worldvolume coordinates $\xi^\rho$,
  \item writing everything only in terms of flat components
  both on the worldvolume and in the target space.
\end{enumerate}
This program is achieved by introducing an auxiliary $0$--form
field $\Pi^{\underline{a}}_i(\xi)$ with an index $\underline{a}$
running in the vector representation of $\mathrm{SO(1,D-1)}$ and a
second index $i$ running in the vector representation of
$\mathrm{SO(1,d-1)}$ and writing the action:
\begin{eqnarray}
  \mathcal{A}^{kin}[d] & = &  \int_{w_d} \, \left [ \, \Pi^{\underline{a}}_j
  \,
  V^{\underline{b}}\, \eta_{\underline{ab}} \, \wedge  \eta^{ji_1} \, e^{i_2} \, \wedge \, \dots
  \, \wedge \, e^{i_d} \, \epsilon_{i_1 \dots i_d} \right.\nonumber\\
  &&\left. - \frac {1}{2d} \,\left (   \, \Pi^{\underline{a}}_i \,
  \Pi^{\underline{b}}_j \, \eta^{ij} \, \eta_{\underline{ab}} \, + \,
  d-2 \, \right)  \, e^{i_1} \, \wedge \, \dots \, \wedge \,
  e^{i_d} \, \epsilon_{i_1 \dots i_d} \right ]
\label{kin}
\end{eqnarray}
The variation of (\ref{kin}) with respect to $\delta
\Pi^{\underline{a}}_j$ yields an equation that admits the unique
algebraic solution:
\begin{equation}
  V^{\underline{a}} \vert_{w_d} = \Pi^{\underline{a}}_i \,
  e^i
\label{urbano}
\end{equation}
Hence the $0$-form $\Pi^{a}_i$ is identified with the
intrinsic components along the world--volume vielbein $e^i$ of the
the bulk vielbein $V^{\underline{a}}$ pulled-back onto the world
volume. In other words the field $\Pi^{a}_i$ is identified by its own
field equation with the field $V^{\underline{a}}_i$ defined in eq.
(\ref{pullobacco}).
On the other hand with the chosen numerical coefficients the
variation of (\ref{kin}) with respect to the world--volume vielbein
$\delta e^i$ yields another equation with the unique algebraic
solution:
\begin{equation}
  \Pi^{\underline{a}}_i \, \Pi^{\underline{b}}_j \, \eta_{\underline{ab}} = \eta_{ij}
\label{flattopullo}
\end{equation}
which is the flat index transcription of eq.(\ref{identh})
identifying the world--volume metric with the pull-back of the bulk
metric. Hence eliminating all the auxiliary fields via their own
equation of motion the first order action (\ref{kin}) becomes
proportional to the second order Nambu--Goto action.
The first order form (\ref{kin}) of the kinetic action is the best
suited one to discuss $\kappa$--symmetry. We consider the case of the supermembrane
\subsection{$\kappa$--symmetry}
\label{kasusy}
In the case of the supermembrane in eleven dimensions the world--volume
is three dimensional and the complete action is simply given by the
kinetic action (\ref{kin}) with $d=3$ plus the Wess-Zumino term,
namely the integral of the $3$--form gauge field $A^{[3]}$.
Explicitly we have:
\begin{equation}
  \mathcal{A}_{M2} = \mathcal{A}^{kin}[d=3] \, - \, {\bf q} \,\int_{w_3} \, A^{[3]}
\label{M2action}
\end{equation}
where ${\bf q}=\pm 1$ is the charge of the supermembrane. As explained in
\cite{Dall'Agata:1999wz}, the background fields, namely the bulk elfbein
$V^{\underline{a}}$ an the bulk three--form $A^{[3]}$ are superspace
differential forms which are assumed to satisfy the Bianchi consistent
rheonomic parametrizations of $D=11$ supergravity as originally given in
\cite{Fdorinal,comments}. Hence, although implicitly, the
action functional (\ref{M2action}) depends both on $11$ bosonic
fields, namely the $X^{\underline{\mu}}(\xi)$ coordinates of bulk
space--time, and on $32$ fermionic fields
$\theta^{\underline{\alpha}}(\xi)$, forming an $11$--dimensional Majorana
spinor. A supersymmetry variation of the background fields is
determined by the rheonomic parametrization of the curvatures and has
the following explicit form:
\begin{eqnarray}
\delta \, V^{\underline{a}} &=& \mbox{\rm i} {\bar \epsilon} \, \Gamma^{\underline{a}} \,
\Psi, \nonumber\\
\label{ordsusy}
\delta \,\Psi &=& {\cal D} { \epsilon} \, -
 \frac{\rm i}{3} \left( \Gamma^{\underline{b}_1\underline{b}_2\underline{b}_3} \,
F_{\underline{a}\underline{b}_1\underline{b}_2\underline{b}_3} - \frac{1}{8}
\Gamma_{\underline{a}\underline{b}_1\dots\underline{b}_4} \,
F^{\underline{b}_1\dots\underline{b}_4} \, \right) \, \epsilon\, V^{\underline{a}}, \\
\delta \, A^{[3]} & = & -  {\bar \epsilon} \, \Gamma^{\underline{a}\underline{b}}
\Psi \, \wedge \, V_{\underline{a}} \, \wedge \, V_{\underline{b}}
\label{d11susies}
\end{eqnarray}
where $\Psi$ is the gravitino $1$--form,
$F_{\underline{a_1,\dots,a_4}}$ are the intrinsic components of the
$A^{[3]}$ curvature and $\epsilon$ is a $32$--component spinor
parameter. Essentially a supersymmetry transformation is a translation
of the fermionic coordinates $\theta \mapsto \theta + \epsilon$.
With such an information the $\kappa$--symmetry invariance of the
action (\ref{M2action}) can be established through a two--line
computation, using the so called $1.5$--order formalism. Technically
this consists of the following: in the action (\ref{M2action}) we vary only the background
fields $V^{\underline a}$ and $A^{[3]}$ with respect to the supersymmetry
transformations (\ref{d11susies}) and, after variation, we use the
first order field equations (\ref{urbano}) and (\ref{flattopullo}). The
action is supersymmetric if all terms that are proportional to
the gravitino $1$--form $\Psi$ cancel against each other. This
does not happen for a generic $32$--component spinor $\epsilon$,
but it does if the latter is of the form:
\begin{eqnarray}
\epsilon &=& \frac{1}{2} \left( 1 +{\rm i} {\bf q} { \widehat{\mathbf{\Gamma}}}
\right)\, \kappa, \nonumber \\
{ \widehat{\mathbf{\Gamma}}} &\equiv & \frac{\epsilon^{ijk}}{3! } \mathbf{\Gamma} _{ijk} =
\frac{\epsilon^{ijk}}{3! } \Pi_i{}^{\underline{a}}  \Pi_j{}^{\underline{b}}
\Pi_k{}^{\underline{c}} \Gamma_{\underline{a}\underline{b}\underline{c}},
\label{projector}
\end{eqnarray}
for a generic spinor $\kappa$. Eq.(\ref{projector}) corresponds to
projection (\ref{projparam}) which halves the spinor
components. It follows that of the $32$ fermionic degrees of freedom
$16$ can be gauged away by $\kappa$--symmetry. The remaining
$16$ are further reduced to $8$ by their field equations.
As one sees, once the supermembrane action is cast into the first order form
(\ref{M2action}),
$\kappa$--symmetry invariance can be implemented in an extremely
simple and elegant way that requires only a couple of algebraic
manipulations with gamma matrices.
\subsection{Extension to $\mathrm{GL(3,\mathbb{R})}$ invariance}
\label{gl3inva}
The classical action (\ref{M2action}) possesses the following invariances:
\begin{enumerate}
  \item $3d$ - diffeomorphism invariance since it is written in
  terms of differential forms and exterior products.
  \item Local $\mathrm{SO(1,2)}$ Lorentz invariance.
  \item $\kappa$--symmetry invariance as described above.
\end{enumerate}
It was noted in paper \cite{Fre:2002kc} that the $\mathrm{SO(1,d-1)}$
invariance of the classical $p$-brane actions can be promoted to
a larger $\mathrm{GL(d,\mathbb{R})}$ invariance by introducing an
additional auxiliary symmetric field $h_{ij}$
In this way one retrieves a supermembrane lagrangian in a
set-up where the $\kappa$--symmetry is easily derived
in an arbitrary supergravity background, avoiding all of its complicacies inherent
to the second order formalism.
\par
The $\mathrm{GL(3,\mathbb{R})}$-covariant classical action which
replaces eq.(\ref{M2action}) can be written as follows:
\begin{eqnarray}
\mathcal{A}_{class} & = & \mathcal{A}_{kin} \, + \, \mathcal{A}_{WZ} \nonumber\\
\mathcal{A}_{kin} & = & \int \, \left\{ \Pi_i^{\underline{a}} \, V^{\underline{b}} \, h^{ij} \, \eta_{\underline{ab}} \,\wedge \, e^j \, \wedge \, e^k \,
\epsilon_{ijk} \right.\nonumber\\
\null & \null & -\ft 16 \,\left. \left [ \Pi^\ell_{\underline{a}} \, \Pi^m_{\underline{b}} \,\eta^{\underline{ab}}
\, h_{\ell m}\, + \, \left( \mbox{det} \, {h}\,\right)  \, \right ] \, e^i \, \wedge \, e^j \, \wedge \, e^k
\, \epsilon_{ijk} \, \right \} \, \nonumber\\
\mathcal{A}_{WZ} & = & - \mathbf{q} \, \int \, A^{[3]} \quad ; \quad \mathbf{q}=\pm 1\nonumber\\
\label{classaction}
\end{eqnarray}
where the choice of the charge sign $\mathbf{q}=\pm 1$ corresponds to
the brane/antibrane respectively.
\par
In equation (\ref{classaction}) the world-volume flat indices $i,j,k$
are raised and lowered with the flat metric $\eta_{ij} =
\mbox{diag}(+,-,-)$ while the Lorentz target space indices, spanning
the vector representation of $\mathrm{SO(1,10)}$ are raised and
lowered with $\eta_{\underline{ab}}= \mbox{diag}\left( +,-,\dots , -\right)
$.
\par
The $\mathrm{GL(3,\mathbb{R})}$ symmetry is realized as follows.
$\forall \, K  \in  \mathrm{GL(3,\mathbb{R})}$, namely for all
non-degenerate $3 \times 3$ matrix $K^{i}_{\phantom{i}j}$, the
following transformations:
\begin{equation}
  \begin{array}{rcl}
    e^i & \mapsto & K^{i}_{\phantom{i}j} \, e^j \\
    \Pi_i^{\underline{a}} & \mapsto & \Pi_\ell^{\underline{a}}  \, \left(  K^{-1} \right) ^{\ell}_{\phantom{\ell}i}  \\
   h^{ij} \, & \mapsto & K^{i}_{\phantom{i}\ell} \, K^{j}_{\phantom{i}m} \, h^{\ell m} \, \left( \mbox{det} K \right) ^{-1}\
  \end{array}
\label{gl3transfa}
\end{equation}
leave the action (\ref{classaction}) invariant. The transition to the
second order formalism is achieved through the implementation of the
field equations for the first order fields, namely $\Pi_j^{\underline{a}}, h_{ij}$ and the dreibein $e^i$.
Let us discuss these equations one by one.
\begin{description}
  \item[$\delta \, \Pi_\ell^{\underline{a}}\, $-- eq.)]  Setting to zero
  this variation of the action (\ref{classaction}) we obtain:
\begin{equation}
h^{\ell i}\left(   V^{\underline{a}} \, \wedge \, e^j \, \wedge \, e^k \,
\epsilon_{ijk} \, - \, \ft 13  \, \Pi^{\underline{a}}_i \, \mbox{Vol}(3) \,\right)  = \, 0
\label{piequa}
\end{equation}
where, by definition, $\mbox{Vol}(3)= \epsilon_{\ell_1\ell_2\ell_3} \,e^{\ell_1} \,
\wedge \, e^{\ell_2} \, \wedge \, e^{\ell_3}$. Eq. (\ref{piequa})
immediately implies:
\begin{equation}
  V^{\underline{a}} =\Pi^{\underline{a}}_{i} \, e^i
\label{Vaei}
\end{equation}
so that the auxiliary fields $\Pi^{\underline{a}}_i $ are interpreted, once on
shell, as the components of the pull-back of the bulk supergravity
vielbein $V^{\underline{a}}$ onto the worldvolume of the $\mathrm{M2}$-brane.
 \item[$\delta {h}^{ij}\,$-- eq.)] Let us define the following $3
 \times 3$ matrix:
\begin{equation}
  \gamma_{ij} \, \equiv \, \left( \, \widehat{{\gamma}} \,\right) _{ij} \, \equiv \,
  \Pi^{\underline{a}}_i \, \Pi^{\underline{b}}_j \, \eta_{\underline{ab}}
\label{gammatra}
\end{equation}
In terms of this matrix the considered variation yields the following matrix equation:
\begin{equation}
  \widehat{\gamma} \, = \, \widehat{h}^{-1} \,\left(  \mbox{det}\, \widehat{h}
  \right)
\label{gammaequa}
\end{equation}
where we have used the standard convention:
\begin{equation}
h_{ij} \, \equiv  \, \left(  \widehat{h}^{-1} \right) _{ij} \quad ;
\quad h^{ij} \, \equiv \, \left ( \widehat{h} \right)^{ij} \quad
\Rightarrow \quad h_{i\ell} \, h^{\ell j} = \delta^j_i
\label{convenziona}
\end{equation}
Eq.(\ref{gammaequa}) admits the unique solution:
\begin{equation}
  \widehat{h} \, = \, \widehat{\gamma}^{-1} \,\left(  \mbox{det}\, \widehat{\gamma} \right)^{1/2}
\label{hingamma}
\end{equation}
The interpretation of these equations is quite obvious. On shell the
matrix $\widehat{\gamma}$ is the pull-back of the bulk metric onto
the $\mathrm{M2}$ worldvolume, written in flat components with respect to a
fiducial dreibein $e^i$. The auxiliary field $h^{ij}$ is just the
inverse of this metric, rescaled by the square root of its
determinant.
  \item[$\delta e^k\,$-- eq.)] The variation of the classical action
  (\ref{classaction}) with respect to the dreibein $\delta e^k$ yields the
  following $2$-form equation:
\begin{equation}
2 \,  h^{i\ell} \Pi_\ell^{\underline{a}} \, V^{\underline{a}} \, \wedge \, e^j
  \,\epsilon_{ijk}\, - \, \ft 12 \left[ \mbox{Tr} \, \left(\widehat{\gamma} \,\widehat{ h} \right) \, +
  \, \left( \mbox{det} \,\widehat{ h} \right) \right] \, e^i \, \wedge \,
  e^j \,  \,\epsilon_{ijk}\, \, = \, 0
\label{eiEqua}
\end{equation}
which is immediately translated into the following matrix equation:
\begin{equation}
  - \, 2 \, \widehat{\gamma} \, \widehat{h} \, + \, \mathbf{1} \, \left[ \mbox{Tr}(\widehat{h} \,
  \widehat{\gamma} ) \,  - \, \left( \mbox{det} \,\widehat{ h} \right) \right]  \, = \, 0
\label{matraequaperdelek}
\end{equation}
If we insert the solution (\ref{hingamma}) for $\widehat{h}$ in terms of
$\widehat{\gamma}$ into eq.(\ref{matraequaperdelek}) we find that it
is identically satisfied. This means that in this formulation the
dreibein equation imposes no new constraints once those for $\Pi$ and
$h$ have been implemented. Such a feature was already stressed in the
original paper \cite{Fre:2002kc}.
\end{description}
The $\mathrm{GL(3,\mathbb{R})}$ invariance of the classical action
can now be used to impose suitable gauges. For instance we always
have enough $\mathrm{GL(3,\mathbb{R})}$ parameters to impose:
\begin{equation}
  \widehat{h} \, = \, \mathbf{\eta} \, \Leftrightarrow \, \widehat{\gamma} \, = \, \mathbf{\eta}
\label{fortegauge}
\end{equation}
In the gauge (\ref{fortegauge}) eq.(\ref{Vaei}) reduces to
eq.(\ref{urbano}) and we retrieve the first order formulation of the classical action without
$\mathrm{GL(3,\mathbb{R})}$ invariance. The second order action is in any case the
same, irrespectively whether we start from the old or from the new first order
formalism.
\par
It is now our program to perform a BRST quantization of the above
classical supermembrane action in presence of constrained ghost fields
(pure spinors). This involves three steps. First we ought to discuss
the relevant BRST algebra, secondly we have to introduce a suitable
gauge fixing term, thirdly we have to verify the BRST invariance of
the complete quantum action and to check the nilpotency of the BRST
operator. In the next section we turn to consideration of the first
of these three steps.
\section{BRST Quantization of $D=11$ supergravity}
\label{BRSgen}
The starting point for the covariant quantization of the supermembrane
with pure spinors is the BRST-quantization of supergravity itself.
Indeed the general strategy of the Berkovits approach consists of
\textit{constraining} some of the \textit{ghost fields} in order to \textit{relax} some of the
\textit{gauge degrees of freedom}. Hence as a preliminary step we
have to write the complete BRST algebra of supergravity
theory which includes the ghosts for all the relevant symmetries,
namely:
\begin{enumerate}
  \item $D=11$ diffeomorphisms
  \item $\mathrm{SO(1,10)}$ Lorentz rotations
  \item $32$--component local supersymmetries
  \item gauge transformations of the $\mathbf{A^{[3]}}$ form.
\end{enumerate}
Successively we look for a consistent set of constraints on the ghost
fields which includes the complete relaxation of diffeomorphisms.
This set of constraints leads to pure spinor constraints on the local supersymmetry
parameters.
\par
So let us start with the first step by deriving the BRST
algebra of $D=11$ supergravity.
We follow a general procedure which was developed
in \cite{Anselmi:1992tj} by extending ideas originally introduced
in \cite{Baulieu:1986dp}.

\subsection{Rheonomy, ghost-forms and 11d supergravity}
\label{ghosteforme}
The main idea of \cite{Anselmi:1992tj} was the extension of  super differential forms
to generalized ghost-form (these are the generalized forms
obtained from a $n$-form by adding a set of ghost-forms with ghost number
$p$ and form degree $n-p$ where $p=1,\dots,n$:
\begin{equation}
  \Omega^{[n]} \rightarrow \sum_{p=0}^n \Omega^{[n-p,p]}\,, \quad \Omega^{[n]} \equiv \Omega^{[n,0]},
\label{ghostiformi}
\end{equation}
the original $n$-form has ghost number zero) starting from the \textit{unique rheonomic parametrization
} of superspace curvatures which defines classical supergravity theory as reviewed
in the introduction. We can condensate the approach of \cite{Anselmi:1992tj} into the principle:
\begin{definizione}\label{pippo}
$\null$\\
The correct BRST algebra is provided by replacing, in the rheonomic parametrization, of the classical
supergravity curvatures each
differential form with its extended ghost-form counterpart while keeping the curvature components
untouched.
Thus one obtains the rheonomic parametrization of the ghost--extended curvatures, whose formal definition
is identical with that of the classical curvatures upon the replacements:
\begin{equation}
  \begin{array}{ccc}
    d & \mapsto & d+\mathcal{S} \\
    \Omega^{[n]} & \mapsto & \sum_{p=0}^n \, \Omega^{[n-p,p]} \
  \end{array}
\label{orletto}
\end{equation}
\end{definizione}
\par
In 11d supergravity,
the supersymmetry transformations (\ref{ordsusy}) which we used to
define the $\kappa$--symmetry of the action (\ref{M2action}) are
just the consequence of the rheonomic parametrizations of the
curvatures for the underlying algebraic structure of $D=11$
supergravity. This latter is not an ordinary Lie superalgebra rather
it is a Free Differential Algebra (see Appendix A for further informations). This means that the list of
generators of the algebra includes, besides a set of $1$--forms, spanning the dual of an ordinary Lie
superalgebra $\mathbb{G}$, also some higher degree forms.
In the specific case of  $D=11$ supergravity  $\mathbb{G}$ is just the $D=11$ Poincar\'e
superalgebra spanned by the following $1$--forms:
\begin{enumerate}
  \item the vielbein $V^a$
  \item the spin connection $\omega^{ab}$
  \item the gravitino $\Psi$
\end{enumerate}
In its minimal formulation suitable to describe the M2 brane,
the relevant FDA includes just one  higher degree generator namely:
\begin{itemize}
  \item the bosonic $3$--form $\mathbf{A^{[3]}}$
\end{itemize}
The complete set of curvatures describing the FDA structure is given below (\cite{Fdorinal,comments}):
\begin{eqnarray}
T^{a} & = & \mathcal{D}V^a - {\rm i} \ft 12 \, \overline{\Psi} \, \wedge \, \Gamma^a \, \Psi \nonumber\\
R^{ab} & = & d\omega^{ab} - \omega^{ac} \, \wedge \, \omega^{cb}
\nonumber\\
\rho & = & \mathcal{D}\Psi \equiv d \Psi - \ft 14 \, \omega^{ab} \, \wedge \, \Gamma_{ab} \, \Psi\nonumber\\
\mathbf{F^{[4]}} & = & d\mathbf{A^{[3]}} - \ft 12\, \overline{\Psi} \, \wedge \, \Gamma_{ab} \, \Psi \,
\wedge \, V^a \wedge V^b
\label{FDAcompleta}
\end{eqnarray}
>From their very definition, by taking a further exterior derivative
one obtains the Bianchi identities, which for brevity we do not
explicitly write (see \cite{comments}). The dynamical theory is
defined, according to a general constructive scheme of supersymmetric theories, by the principle
of rheonomy (see \cite{castdauriafre} ) implemented into Bianchi identities.
Indeed there is a unique rheonomic parametrization of the curvatures (\ref{FDAcompleta}) which solves the
Bianchi identities and it  is the following one:
\begin{eqnarray}
T^a & = & 0 \nonumber\\
\mathbf{F^{[4]}} & = & F_{a_1\dots a_4} \, V^{a_1} \, \wedge \dots \wedge \, V^{a_4} \nonumber\\
\rho & = & \rho_{a_1a_2} \,V^{a_1} \, \wedge \, V^{a_2} - {\rm i} \ft 12 \,
\left(\Gamma^{a_1a_2 a_3} \Psi \, \wedge \, V^{a_4} + \ft 1 8
\Gamma^{a_1\dots a_4 m}\, \Psi \, \wedge \, V^m
\right) \, F^{a_1 \dots a_4} \nonumber\\
R^{ab} & = & R^{ab}_{\phantom{ab}cd} \, V^c \, \wedge \, V^d
+ {\rm i} \, \rho_{mn} \, \left( \ft 12 \Gamma^{abmn} - \ft 2 9 \Gamma^{mn[a}\, \delta^{b]c} + 2 \,
\Gamma^{ab[m} \, \delta^{n]c}\right) \, \Psi \wedge V^c\nonumber\\
 & &+\overline{\Psi} \wedge \, \Gamma^{mn} \, \Psi \, F^{mnab} + \ft 1{24} \overline{\Psi} \wedge \,
 \Gamma^{abc_1 \dots c_4} \, \Psi \, F^{c_1 \dots c_4}
\label{rheoFDA}
\end{eqnarray}
The expressions (\ref{rheoFDA}) satisfy the Bianchi identities provided the space--time components
of the curvatures satisfy the following constraints
\begin{eqnarray}
0 & = & \mathcal{D}_m F^{mc_1 c_2 c_3} \, + \, \ft 1{96} \, \epsilon^{c_1c_2c_3 a_1 a_8} \, F_{a_1 \dots a_4}
\, F_{a_5 \dots a_8}  \nonumber\\
0 & = & \Gamma^{abc} \, \rho_{bc} \nonumber\\
R^{am}_{\phantom{bm}cm} & = & 6 \, F^{ac_1c_2c_3} \,F^{bc_1c_2c_3} -
\, \ft 12 \, \delta^a_b \, F^{c_1 \dots c_4} \,F^{c_1 \dots c_4}
\label{fieldeque}
\end{eqnarray}
which are the  space--time field equations.
\par
At this stage the extension to ghost-forms becomes very easy. To the
$(0,1)$--component of each element of the cotangent  basis we give a
specific name which will be useful for its later interpretation in
the covariant quantization of the supermembrane. Explicitly we set
\begin{equation}
  \begin{array}{rcl}
    V^a & \Rightarrow & V^a + \xi^a \\
    \Psi & \Rightarrow & \Psi + \lambda \\
    \omega^{ab}  & \Rightarrow & \omega^{ab} + \epsilon ^{ab}\\
    A^{[3]} & \Rightarrow & A^{[3]} + \sum_{i=1}^{3} \, c^{[3-i,i]} \\
\end{array}
\label{ghosti}
\end{equation}
where $c^{[3-i,i]}$ are $3-i$--forms of ghost number $i$.
\par
By implementing principle \ref{pippo} and using both the definition
(\ref{FDAcompleta}) and the classical rheonomic parametrization of
the FDA curvatures (\ref{rheoFDA}) we obtain the BRST algebra, namely
the BRST transformations of all the ghost and physical fields.
Explicitly, in the highest ghost number sector we find:
\begin{equation}
  \begin{array}{rcl}
    s \, \xi ^a \, - \, \epsilon ^{ab} \, \xi_{b} \, & = & \, \ft {\rm i} 2 \, \overline{\lambda} \, \Gamma^a \, \lambda
    \\
    s \, \epsilon ^{ab} \,  - \, \epsilon ^{ac} \, \epsilon^{cb} \,  & = & \begin{array}{l}
    \null\\
    \null\\
      R^{ab}_{\phantom{ab}mn} \, \xi^m \, \xi^{n} \, \\
      + \,{\rm i} \, {\bar \rho}_{mn} \, \left( \ft 12 \Gamma^{abmn} - \ft 2 9 \Gamma^{mn[a}\, \delta^{b]c} + 2 \,
\Gamma^{ab[m} \, \delta^{n]c}\right) \,  \lambda \,  \xi^c \\
      \, + \, \overline{\lambda}  \, \Gamma^{mn} \, \lambda \, F^{mnab} + \ft 1{24} \overline{\lambda}  \,
 \Gamma^{abc_1 \dots c_4} \, \lambda \, F^{c_1 \dots c_4} \\
    \end{array} \\
    s \, \lambda \, - \, \ft 1 4 \, \epsilon ^{ab} \, \Gamma_{ab} \,
 \lambda & = & \begin{array}{l}
   \null \\
   \rho_{a_1a_2} \,\xi^{a_1} \,  \xi^{a_2} \\
   - {\rm i} \ft 12 \,
\left(\Gamma^{a_1a_2 a_3} \lambda \, \xi^{a_4} + \ft 1 8
\Gamma^{a_1\dots a_4 m}\, \lambda \,  \xi^m
\right) \, F^{a_1 \dots a_4} \
\end{array} \\
 s \mathbf{c}^{[0,3]} & = & \begin{array}{l}
   \ft 12\, \overline{\lambda} \,  \Gamma_{ab} \, \lambda \,
\xi^a \, \xi^b \,  + \,  F_{a_1\dots a_4} \, \xi^{a_1} \, \dots \, \xi^{a_4} \
 \end{array} \\
   s \mathbf{c}^{[1,2]} \, +\,  d\mathbf{c}^{[0,3]}\, & = & \begin{array}{l}
     \null \\
     \overline{\lambda} \,  \Gamma_{ab} \, \Psi \,
\xi^a \, \xi^b \, + \, \overline{\lambda} \,  \Gamma_{ab} \, \lambda \,
\xi^a \, V^b \, \\
     4 \, F_{a_1\dots a_4} \, \xi^{a_1} \, \dots \,
V^{a_4} \\
   \end{array} \\
   s \mathbf{c}^{[2,1]} \, +\,  d\mathbf{c}^{[1,2]}\, & = & \begin{array}{l}
     \null\\
     \null \\
     \, \overline{\lambda} \,  \Gamma_{ab} \, \lambda \,
V^a \, \wedge \, V^b \, + \, \overline{\Psi} \,  \wedge \, \Gamma_{ab} \, \Psi \,
\xi^a \, \xi^b\, \\
     \, + \,  2 \, \overline{\Psi} \,  \wedge \, \Gamma_{ab} \, \lambda \,
\xi^a \, V^b \\
    \, + \,  6 \, F_{a_1\dots a_4} \, \xi^{a_1} \, \dots \, V^{a_3} \,
\wedge \, V^{a_4} \
   \end{array} \
  \end{array}
\label{highestghost}
\end{equation}
The next bit of information to be extracted from the quantum rheonomic parametrizations are the
the BRST transformations of the physical fields, yet prior to that it
is convenient to introduce a Lorentz covariant formalism by splitting
the ghost extended Lorentz covariant derivative in the following way:
\begin{eqnarray}
\widehat{D} & = & \widehat{d} \, + \,\widehat{ \omega} ^{ab} \, J_{ab} \nonumber\\
\null & = & d \, + \, s \, + \, \omega^{ab}\, J_{ab} \,+ \, \epsilon^{ab}\, J_{ab} \nonumber\\
\null & = & \mathcal{D} \, + \,  \mathcal{S}\nonumber\\
\mbox{where} & \null & \null \nonumber\\
\mathcal{D} & = & d \, +\, \omega^{ab}\, J_{ab} \quad \mbox{Lorentz covariant external
derivative}\nonumber\\
\mathcal{S} & = & s \, + \,\epsilon^{ab}\, J_{ab}\quad \mbox{Lorentz covariant
BRST variation}
\label{Lorenzocovari}
\end{eqnarray}
where $J_{ab}$ denotes the standard generators of the $\mathrm{SO(1,10)}$ Lie
algebra.
\par
The quantum rheonomic parametrization of the curvatures implies that
the above operators satisfy the following algebra:
\begin{equation}
  \begin{array}{rcl}
    \mathcal{S}^2 & = & \begin{array}{l}
    \null\\
    \null\\
     \big [ R^{ab}_{\phantom{ab}mn} \, \xi^m \, \xi^{n} \, \\
      + \,{\rm i} \, {\bar \rho}_{mn} \, \left( \ft 12 \Gamma^{abmn} - \ft 2 9 \Gamma^{mn[a}\, \delta^{b]c} + 2 \,
\Gamma^{ab[m} \, \delta^{n]c}\right) \,  \lambda \,  \xi^c \\
      \, + \, \overline{\lambda}  \, \Gamma^{mn} \, \lambda \, F^{mnab} + \ft 1{24} \overline{\lambda}  \,
 \Gamma^{abc_1 \dots c_4} \, \lambda \, F^{c_1 \dots c_4} \big ] \, J_{ab}\\
    \end{array} \\
\mathcal{D}^2 & = & \begin{array}{l}
    \null\\
    \null\\
     \big [ R^{ab}_{\phantom{ab}mn} \, V^m \, \wedge V^{n} \, \\
      + \,{\rm i} \, {\bar \rho}_{mn} \, \left( \ft 12 \Gamma^{abmn} - \ft 2 9 \Gamma^{mn[a}\, \delta^{b]c} + 2 \,
\Gamma^{ab[m} \, \delta^{n]c}\right) \,  \Psi \,  \wedge \, V^c \\
      \, + \, \overline{\Psi}  \, \wedge \, \Gamma^{mn} \, \Psi \, F^{mnab} + \ft 1{24} \overline{\Psi}
      \, \wedge \,
 \Gamma^{abc_1 \dots c_4} \, \Psi \, F^{c_1 \dots c_4} \big ] \, J_{ab}\\
    \end{array} \\
    \mathcal{S}\, \mathcal{D} \, + \,  \mathcal{D}\, \mathcal{S} \, & = & \begin{array}{l}\null\\
    \null\\
    \null\\
     \big [2 \, R^{ab}_{\phantom{ab}mn} \, V^m \, \xi^{n} \, \\
      + \,{\rm i} \, {\bar \rho}_{mn} \, \left( \ft 12 \Gamma^{abmn} - \ft 2 9 \Gamma^{mn[a}\, \delta^{b]c} + 2 \,
\Gamma^{ab[m} \, \delta^{n]c}\right) \,  \lambda   \wedge \, V^c \\
+ \,{\rm i} \, {\bar \rho}_{mn} \, \left( \ft 12 \Gamma^{abmn} - \ft 2 9 \Gamma^{mn[a}\, \delta^{b]c} + 2 \,
\Gamma^{ab[m} \, \delta^{n]c}\right) \,  \Psi   \wedge \, \xi^c \\
      + \, 2 \, \overline{\lambda}  \, \Gamma^{mn} \, \Psi \, F^{mnab} + \ft 1{12} \overline{\lambda}
      \,
 \Gamma^{abc_1 \dots c_4} \, \Psi \, F^{c_1 \dots c_4} \big ] \, J_{ab} \
  \end{array}\\
  \end{array}
\label{algebrona}
\end{equation}
Utilizing these Lorentz covariant operators we can now write the BRST
transformation of the physical supergravity fields as they follow
from the quantum rheonomic parametrizations of the ghost-extended
curvatures. We find:
\begin{equation}
  \begin{array}{ccl}
    {\mathcal{S}}\,  V^a \, & = & - \, \mathcal{D}\, \xi^a \, + \, {\rm i} \, \overline{\Psi} \, \Gamma^a \, \lambda \, \\
    {\mathcal{S}} \,\Psi \, & = & - \, \mathcal{D}\, \lambda \, + \, 2 \, \rho_{a_1a_2} \,V^{a_1} \, \wedge \, \xi^{a_2}\\
   \null & \null & - {\rm i} \ft 12 \,
\left(\Gamma^{a_1a_2 a_3} \lambda \,  V^{a_4} + \ft 1 8
\Gamma^{a_1\dots a_4 m}\, \lambda \,  V^m
\right) \, F^{a_1 \dots a_4} \\
    \null & \null & - {\rm i} \ft 12 \,
\left(\Gamma^{a_1a_2 a_3} \Psi \, \xi^{a_4} + \ft 1 8
\Gamma^{a_1\dots a_4 m}\, \Psi \,  \xi^m
\right) \, F^{a_1 \dots a_4} \\
  \mathcal{S}\omega^{ab} & = & - \, \mathcal{D}\epsilon^{ab} \, + \,  2 \, R^{ab}_{\phantom{ab}mn} \, V^m \, \xi^{n} \, \\
    \null & \null &
      + \,{\rm i} \, {\bar \rho}_{mn} \, \left( \ft 12 \Gamma^{abmn} - \ft 2 9 \Gamma^{mn[a}\, \delta^{b]c} + 2 \,
\Gamma^{ab[m} \, \delta^{n]c}\right) \,  \lambda    \, V^c \\
    \null & \null & + \,{\rm i} \, {\bar \rho}_{mn} \, \left( \ft 12 \Gamma^{abmn} - \ft 2 9 \Gamma^{mn[a}\, \delta^{b]c} + 2 \,
\Gamma^{ab[m} \, \delta^{n]c}\right) \,  \Psi    \, \xi^c  \\
    \null & \null & + \,  2 \, \overline{\lambda}  \, \Gamma^{mn} \, \Psi \, F^{mnab} + \ft 1{12} \overline{\lambda}
      \,
 \Gamma^{abc_1 \dots c_4} \, \Psi \, F^{c_1 \dots c_4} \\
 \mathcal{S }\mathbf{A^{[3]}} & = & - \, d \mathbf{c^{[2,1]}} \, +  \, \overline{\Psi} \, \wedge \, \Gamma_{ab} \, \lambda \,
\wedge \, V^a \wedge V^b \nonumber\\
\null & \null & +  \, \overline{\Psi} \, \wedge \, \Gamma_{ab} \, \Psi \,
\wedge \, V^a \wedge \xi^b \, + \,  4 \, F_{a_1\dots a_4} \, V^{a_1} \, \dots \,
\xi^{a_4} \
  \end{array}
\label{phystransfor}
\end{equation}
This concludes our presentation of the unconstrained BRST algebra of
$D=11$ supergravity. In the next section we discuss its constrained
version.
\subsection{Constrained BRST algebra}
\label{constBRS}
In agreement with the general philosophy of the pure spinor approach we
reconsider the BRST algebra presented in the previous section setting to zero
the parameters of diffeomorphisms, namely:
\begin{equation}
  \xi^{a} \simeq 0
\label{dffeoliberi}
\end{equation}
If we set such a  constraint,  the BRST transformation algebra on the
ghosts is easily read off from eq.s (\ref{highestghost}) that become:
\begin{eqnarray}
 0 & = & \overline{\lambda} \, \Gamma^a \, \lambda\nonumber\\
  \mathcal{S} \epsilon ^{ab} &  = & \,  \overline{\lambda}  \, \Gamma^{mn} \, \lambda \, F^{mnab} + \ft 1{24} \overline{\lambda}  \,
 \Gamma^{abc_1 \dots c_4} \, \lambda \, F^{c_1 \dots c_4} \nonumber\\
\mathcal{S} \lambda &  = & \,0 \nonumber\\
\mathcal{S} \mathbf{c}^{[2,1]} & = & \overline{\lambda}  \, \Gamma^{mn} \, \lambda
\, V_m \, \wedge \, V_n
\label{hatalgebra}
\end{eqnarray}
At the same time when $\xi^{a} \simeq 0$ the BRST transformations of
the physical fields (\ref{phystransfor}) become:
\begin{eqnarray}
\mathcal{S}  V^a &   = & {\rm i} \, \overline{\Psi} \, \Gamma^a \, \lambda \, \label{SdiVa}
\nonumber \\
\mathcal{S}  \omega^{ab} &  = &  - \, \mathcal{D}\epsilon^{ab} \,
      + \,{\rm i} \, {\bar \rho}_{mn} \, \left( \ft 12 \Gamma^{abmn} - \ft 2 9 \Gamma^{mn[a}\, \delta^{b]c} + 2 \,
\Gamma^{ab[m} \, \delta^{n]c}\right) \,  \lambda    \, V^c
\nonumber \\
    \null & \null & + \,  2 \, \overline{\lambda}  \, \Gamma^{mn} \, \Psi \, F^{mnab} + \ft 1{12} \overline{\lambda}
      \,
 \Gamma^{abc_1 \dots c_4} \, \Psi \, F^{c_1 \dots c_4}\label{Sdiomega}
 \nonumber \\
 \mathcal{S} \,\Psi &  = & - \, \, \mathcal{D}\, \lambda \, - \, {\rm i} \ft 12 \,
\left(\Gamma^{a_1a_2 a_3} \lambda \,  V^{a_4} + \ft 1 8
\Gamma^{a_1\dots a_4 m}\, \lambda \,  V^m
\right) \, F^{a_1 \dots a_4}\nonumber\\
\null & \equiv & - \nabla \, \lambda \label{Sdipsi}
\nonumber
\\
 \mathcal{S} \, \mathbf{A^{[3]}}  & = & - \, d \mathbf{c^{[2,1]}} \, +  \, \overline{\Psi} \, \wedge \,
 \Gamma_{ab} \, \lambda \,
\wedge \, V^a \wedge V^b
\label{phystransfor2}
\end{eqnarray}
\par

\begin{description}
  \item[a)] The first equation in (\ref{hatalgebra}) is the pure
  spinor constraint of the $11$-dimensional superparticle.
  \item[b)] From the second of eq.s (\ref{hatalgebra}) we deduce
  that for flat space there are no further constraints by removing also the
  Lorentz ghosts
\item[c)] The last equation in (\ref{hatalgebra}) shows that if one
removes also the $c^{[2,1]}$ ghost, namely that associated with the
three form gauge transformation, one more constraint on the spinor
lambda pops up. Such a constraint was  already seen in the literature
\cite{Berkovits:2002uc}. Indeed if we want to reinstall the gauge
invariance of the three form we have to set $c^{[2,1]} \approx 0$ and
closure of the BRST operator implies:
\begin{equation}
  \overline{\lambda} \, \Gamma_{\underline{ab}} \, \lambda \, V^{\underline{a}} \wedge V^{\underline{b}} \, = \, 0
\label{newconstra}
\end{equation}
\end{description}
Let us now consider the pull-back of the constraint
(\ref{newconstra}) on the worldvolume of the $\mathrm{M2}$-brane. Using the
gauge (\ref{fortegauge}) we immediately see that
eq.(\ref{newconstra}) is just equivalent to writing:
\begin{equation}
   \overline{\lambda} \, \Gamma_{\underline{ab}} \, \lambda \,
 \Pi^{\underline{a}}_{[i} \, \Pi^{\underline{b}}_{j]} \,
= \, 0
\label{hornus}
\end{equation}
where we have reinstalled the underlined notation $\underline{a},\underline{b},\underline{c},\dots$
for the $\mathrm{SO(1,10)}$ vector indices, which, in the
discussion of bulk supergravity, we had suppressed since no other type of indices was
needed. Here, coming back to the brane we have to distinguish target
space from worldvolume indices.
\par
It is obvious that the constraint
(\ref{hornus}) is certainly satisfied if we enforce the stronger
one:
\begin{equation}
  \overline{\lambda} \, \Gamma_{\underline{ab}} \, \lambda \,
  \Pi^{\underline{b}}_{j} \,
= \, 0
\label{fortissimo}
\end{equation}
found by Berkovits \cite{Berkovits:2002uc} in his hamiltonian formulation of the
quantum $\mathrm{M2}$ brane \'a la pure spinors in a flat superspace
background.
\par
In the next section we show that the first of eq.s (\ref{hatalgebra}) and
(\ref{fortissimo}) are sufficient to write a Lagrangian for the supermembrane which is
BRST invariant on an arbitrary $D=11$ supergravity background.

\section{The gauge fixing and its  cohomological meaning}
\label{gaugiofisso}

\subsection{Gauge fixing and Dirac equation}
\label{gaugeDirac}

Our next task is constructing an appropriate \textit{gauge
fixing term} to be added in the usual way to the classical action, namely by writing:
\begin{eqnarray}
  \mathcal{A}_{class} & \mapsto & \mathcal{A}_{quantum} \, \equiv \, \mathcal{A}_{class} +
  \mathcal{A}_{GF}\nonumber\\
  \mathcal{A}_{GF} & = & \mathcal{S} \, \int \, \Phi_{gauge}
\label{quantaction}
\end{eqnarray}
where $\mathcal{S}$ denotes the BRST operator discussed in section
\ref{constBRS} and the \textit{gauge fermion} has the standard
structure:
\begin{eqnarray}
  \Phi_{gauge} \, = \, \mbox{antighost}\, \times \, \mbox{gauge fixing}
\label{phigaugo}\nonumber
\end{eqnarray}
Assuming, as it usual, that:
\begin{eqnarray}
  \mathcal{S} \, (\mbox{antighost}) \, = \, \mbox{Lagrange multiplier}
\label{standardo}\nonumber
\end{eqnarray}
the variation in the \textit{Lagrange multiplier} will implement the
\textit{gauge fixing condition} as a field equation of the
BRST invariant \textit{quantum action}.
What is the the local symmetry to be gauge fixed?
At first sight it might seem that this is $\kappa$--symmetry, under
which the classical action is invariant. Unfortunately, despite several
attempts no clear and definitive answers came from
\cite{Matone:2002ft,Grassi:2002xf, Aisaka:2005vn}.

In the present case, the classical action $\mathcal{A}_{class}$ is not
invariant under BRST symmetry (indeed the $\kappa$--symmetry parameter
is replaced by a pure spinor $\lambda$), but the action $\mathcal{A}_{quantum}$
is invariant since the variation of the the first
 term is compensated by the variation of the second term $\mathcal{S} \, \int \, \Phi_{gauge}$.
 This is possible only if the BRST operator is not nilpotent. This seems strange since
 it is exactly the requirement of the nilpotency of the BRST operator that has led to pure spinor constraints.
 Nevertheless,
 we have to recall that the pure spinor constraints are first class constraints and they
 generate a gauge symmetry on the conjugated fields, namely the antighosts.
 Therefore, the BRST operator is nilpotent
 modulo the gauge symmetries on the antighosts. Those symmetries are crucial for the
 construction of the action.
 \par
Hence we look for a gauge fixing term with the following properties:
\begin{enumerate}
  \item It should be $\mathrm{GL(3,\mathbb{R})}$ invariant.
  \item It should not depend neither on the dreibein $e^k$ nor on the
  auxiliary fields $\Pi^{\underline{a}}_i$, and $h_{ij}$, so that it
  will not perturb the field equations of those fields and their
  elimination leading to the second order action.
  \item It should yield a propagation equation for the fermionic
  coordinates $\theta$ of target superspace which we can recognize as a
  standard Dirac equation for spinors on the worldvolume. Notice that
  the equations obtained from the $\mathcal{A}_{class}$ are the equations of motion
  only for half of the $\theta$'s.
  \item Imposing BRST invariance of the action with such a gauge
  fermion should be consistent with the nilpotency of the BRST
  operator up to gauge transformations, namely:
\begin{equation}
  \mathcal{S}^2 \, \mbox{field} \, = \, \mbox{gauge transformation of that field}
\label{nilpotentcirca}
\end{equation}
\item The available gauge transformations, apart from Lorentz
symmetry, must be those generated by the two constraints:
\begin{eqnarray}
0 & \approx & \overline{\lambda} \, \Gamma^{\underline{a}} \,
\lambda \label{costretti_a}\\
0 & \approx & \overline{\lambda} \, \Gamma_{\underline{ab}} \, \lambda \,
  \Pi^{\underline{b}}_{j} \,  \label{costretti_b}
\end{eqnarray}
\end{enumerate}
It is quite remarkable and suggestive that the answer to the above
list of requirements is provided by an essentially unique
gauge-fixing endowed with a profound cohomological meaning.
Before entering its description and
for the reader convenience we have listed in table \ref{allfieldi}  all the fields which enter
our construction, specifying also their grading and representation
assignments under the two local groups $\mathrm{SO(1,2)}$ and
$\mathrm{SO(1,10)}$.
\begin{table}[!tb]
{\small \begin{center}
\begin{tabular}{||l||c|c||c|c||l|}\hline
Field  & Form  & ghost  & $\mathrm{SO(1,2)}$ & $\mathrm{SO(1,10)}$ & Phys.    \\[2mm]
name   & degree & number & repr. & repr. & Role \\
\hline
\hline
$V^{\underline{a}}$ & 1 & 0 & $\mathbf{1}$ & $\mathbf{11}$ & Target
space
Vielbein \\[2mm]
$\Psi$ & 1 & 0 & $\mathbf{1}$ & $\mathbf{32}$ & Target
space
gravitino\\[2mm]
$e^i$ & 1 & 0 & $\mathbf{3}$ & $\mathbf{1}$ & worldvolume
dreibein\\[2mm]
$\Pi^{\underline{a}}_i$ & 0 & 0 & $\mathbf{3}$ & $\mathbf{11}$ & Auxiliary field\\[2mm]
$h_{ij}$ & 0 & 0 & $\underbrace{\mathbf{{3}}\otimes \mathbf{{3}}}_{symm}$ & $\mathbf{1}$ & Auxiliary field\\[2mm]
$\lambda$ & 0 & 1 & $\mathbf{1}$ & $\mathbf{32}$ & susy ghost: pure spinor\\[2mm]
$w$ & 0 & -1 & $\mathbf{1}$ & $\mathbf{32}$ & antighost\\[2mm]
$\Delta$ & 0 & 0 & $\mathbf{1}$ & $\mathbf{32}$ & Lagrange multiplier\\[2mm]
\hline
$\psi_i$ & 0 & 0 & $\mathbf{3}$ & $\mathbf{32}$ & name of the gravitino\\
\null & \null & \null & \null & \null & components along the dreibein\\[2mm]
\hline
\hline
\end{tabular}
\end{center}}
\caption{\footnotesize
{\it List of all the fields, of their gradings and of their representation
assignments.}
All the fields listed above the last line of the table are true
fields appearing in the lagrangian and respect to which we are supposed
to vary the action. Below the line we have listed an object $\psi_i$
which is not a true field, rather it is the name given to the
components of the pull back of the gravitino $\Psi$ when it is
expanded along the dreibein. In some sense $\psi_i$ is the fermionic
counterpart of the auxiliary field $\Pi^{\underline{a}}_i$ but
differently from this latter it does not appear in the action since
the fermionic action is anyhow already of the first order and
geometrical.\label{allfieldi}}
\end{table}
\vskip .3cm
Let us then state that the appropriate gauge fixing is given by the
following $3$--form equation:
\begin{equation}
  \Gamma_{\underline{ab}} \, \Psi \, \wedge \, V^{\underline{a}} \, \wedge \,
  V^{\underline{b}} \, = \, 0
\label{gaugosuso}
\end{equation}
which should hold true upon pull-back on the worldvolume.

As we anticipated the choice (\ref{gaugosuso})  has a deep
cohomological meaning, since the corresponding gauge fixing term
added to the Lagrangian is related to the basic cohomology $4$-cycle
of the super-Poincar\'e algebra responsible for the extension of the
latter to the FDA  of M-theory (see eq.(\ref{FDAcompleta})) and, ultimately, to the very existence of
supermembranes.
\par
According to table \ref{allfieldi} and eq.(\ref{urbano})\footnote{
Implementing the $\mathrm{GL(3,\mathbb{R})}$ gauge (\ref{fortegauge}), or the equation of motion of the dreibein, if we prefer the formulation without
$h^{ij}$ and without $\mathrm{GL(3,\mathbb{R})}$ invariance.} equation (\ref{gaugosuso})
becomes:
\begin{equation}
  \mathbf{\Gamma}_{ij} \, \psi_{k} \, \epsilon^{ijk} \, = \, 0
\label{diraccus}
\end{equation}
where we have introduced the following convenient notation:
\begin{equation}
  \mathbf{\Gamma}_{i_1\dots i_n} = \Gamma_{\underline{a}_1\dots \underline{a}_n
  }\, \Pi^{\underline{a}_1}_{i_1} \, \dots \,
  \Pi^{\underline{a}_n}_{i_n} \quad ; \quad n \le 3 \,
\label{grassegamme}
\end{equation}
If we recall that the gravitino $1$--form $\Psi$ always begins with
the differential of the fermionic superspace coordinate:
\begin{equation}
  \Psi \, = \, d\theta + \mbox{more}
\label{psidtheta}
\end{equation}
it is evident that the constraint (\ref{diraccus}) is a sort of Dirac equation for worldvolume spinors.
\par

\subsection{Gauge fixing and FDA}
\label{gaugeFDA}

The relation with  cohomology and with the structure of the Free
Differential Algebra is provided by the following considerations.
\par
Let us recall the basic Fierz identity responsible for the existence
of the $4$-cycle which extends the $D=11$ super-Poincar\'e Lie algebra
to the FDA (\ref{FDAcompleta}). It is:
\begin{equation}
  \overline{\Psi }\, \wedge \, \Gamma_{\underline{ab}} \, \Psi \, \wedge \,
  \overline{\Psi} \,\wedge \, \Gamma^{\underline{a}} \, \Psi \, \wedge \,
  V^{\underline{b}} \, = \, 0
\label{basicfierzus}
\end{equation}
As it is extensively discussed in the literature
\cite{Fdorinal,comments}, eq.(\ref{basicfierzus}) implies that the
$4$--form:
\begin{equation}
  \Omega^{[4]} \, \equiv \, \overline{\Psi }\, \wedge \, \Gamma_{\underline{ab}} \, \Psi \, \wedge \,
  V^{\underline{a}} \,\wedge \,
  V^{\underline{b}}
\label{4cyclus}
\end{equation}
is closed modulo superPoincar\'e curvatures:
\begin{equation}
  d \Omega^{[4]} = 0 \quad \mbox{at $T^{\underline{a}} = R^{\underline{ab}} = \rho = 0$}
\label{chiusura}
\end{equation}
and this gives origin to the $3$--form $A^{[3]}$ which extends the super-Poincar\'e algebra to an FDA and provides the missing
degrees of freedom of the $D=11$ supergravity multiplet.
Furthermore, since it naturally couples to the worldvolume of a two-dimensional object, $\mathbf{A}^{[3]}$
allows for the existence of the $\mathrm{M2}$ brane.
\par
One can immediately note that the proposed gauge fixing (\ref{gaugosuso}) is simply
the variation in $\delta \overline{\Psi}$ of the $4$--cycle
$\Omega^{[4]}$.
\par
Extending the differential form algebra to the algebra of differential ghost--forms:
\begin{eqnarray}
\Psi & \rightarrow & \mathbf{\Psi} \, \equiv \, \Psi \, + \, \lambda \nonumber\\
V^{\underline{a}} & \mapsto & \mathbf{V}^{\underline{a}} \, = \,
V^{a} \quad \mbox{at $\xi^{\underline{a}} \, \approx \, 0$}
\nonumber\\
d & \mapsto & \mathbf{d} \equiv d \, + \, \mathcal{S}
\label{estesone}
\end{eqnarray}
the $4$--cycle (\ref{4cyclus}) of $d$-cohomology  is immediately promoted to a
$4$--cycle of the $\mathbf{d}$-operator by writing:
\begin{equation}
  \widehat{\Omega}^{[4]} \, = \, \overline{\mathbf{\Psi} }\, \wedge \, \Gamma_{\underline{ab}} \, \mathbf{\Psi} \, \wedge \,
  \mathbf{V}^{\underline{a}} \,\wedge \,
  \mathbf{V}^{\underline{b}} \quad ; \quad \mathbf{d} \,
  \widehat{\Omega}^{[4]} \, = \, 0 \quad\mbox{modulo curvatures}
\label{extendedOmega}
\end{equation}
Expanding in ghost-number, from eq.(\ref{extendedOmega}) we obtain:
\begin{eqnarray}
\widehat{\Omega}^{[4]} & = & \Omega^{[4,0]} \, + \, \Omega^{[3,1]} \, + \, \Omega^{[2,2]}\nonumber\\
\Omega^{[4,0]}  & \equiv & \overline{\Psi }\, \wedge \, \Gamma_{\underline{ab}} \, \Psi \, \wedge \,
  V^{\underline{a}} \,\wedge \,
  V^{\underline{b}}\nonumber\\
  \Omega^{[3,1]}  & \equiv & 2  \, \overline{\lambda }\, \Gamma_{\underline{ab}} \, \Psi \, \wedge \,
  V^{\underline{a}} \,\wedge \,
  V^{\underline{b}}\nonumber\\
  \Omega^{[2,2]}  & \equiv &  \overline{\lambda } \, \Gamma_{\underline{ab}} \, \lambda \,
  V^{\underline{a}} \,\wedge \,
  V^{\underline{b}}\nonumber\\
\label{omegapq}
\end{eqnarray}
and the descent equations:
\begin{eqnarray}
\mathcal{S} \, \Omega^{[p,q-1]} \, + \, d \, \Omega^{[p-1,q]} \, = \,
\Xi^{[p,q]}
\label{descenteque}
\end{eqnarray}
where $\widehat{\Xi}^{[5]}$ is the $5$--form expressing the deviation
from zero of $\mathbf{d}\widehat{\Omega}^{[4]}$ in presence of
curvatures. Explicitly, since the torsion $T^{\underline{a}}$ is always kept zero (see
(\ref{rheoFDA}) ) we have:
\begin{equation}
  \widehat{\Xi}^{[5]} \, = \, - 2 \, \overline{\mathbf{\Psi}} \,
  \Gamma_{\underline{ab}} \, \widehat{\mathbf{\rho}} \, \wedge \,
  \mathbf{V}^{\underline{a}} \, \wedge \, \mathbf{V}^{\underline{b}}
\label{creanza}
\end{equation}
where, in force of the rheonomic parametrizations (\ref{rheoFDA}), we
have:
\begin{eqnarray}
\widehat{\mathbf{\rho}} & = & {\mathbf{\rho}}^{[2,0]} \, + \, {\mathbf{\rho}}^{[1,1]} \nonumber\\
{\mathbf{\rho}}^{[2,0]} & = & \rho_{\underline{ab}} \, V^{\underline{a}} \, \wedge \,
V^{\underline{b}}\nonumber\\
{\mathbf{\rho}}^{[1,1]} & = &- \, {\rm i} \ft 12 \,
\left(\Gamma^{a_1a_2 a_3} \lambda \,  V^{a_4} + \ft 1 8
\Gamma^{a_1\dots a_4 m}\, \lambda \,  V^m
\right) \, F^{a_1 \dots a_4}\nonumber\\
\label{rhoquite}
\end{eqnarray}
Crucial for the gauge symmetries of the quantum action with the gauge fixing (\ref{gaugosuso})
is the case $p=3 \,, \, q=2$ of the
descent equation (\ref{descenteque}), which explicitly reads:
\begin{equation}
  \mathcal{S} \, \left(\,2 \,\overline{\lambda }\, \Gamma_{\underline{ab}} \, \Psi \, \wedge \,
  V^{\underline{a}} \,\wedge \,
  V^{\underline{b}} \right) \, + \, d \, \left(\overline{\lambda } \, \Gamma_{\underline{ab}} \, \lambda \,
  V^{\underline{a}} \,\wedge \,
  V^{\underline{b}} \right) \, = \, - 2 \, \overline{{\lambda}} \,
  \Gamma_{\underline{ab}} \, {\rho}^{[1,1]} \, \wedge \,
  {V}^{\underline{a}} \, \wedge \, {V}^{\underline{b}}
\label{egoista}
\end{equation}
Recalling the definition (\ref{Sdipsi}) of the supercovariant
derivative $\nabla$ and comparing with (\ref{rhoquite}),
eq.(\ref{egoista}) can be rewritten as:
\begin{equation}
  \mathcal{S} \, \left(\,2 \,\overline{\lambda }\, \Gamma_{\underline{ab}} \, \Psi \, \wedge \,
  V^{\underline{a}} \,\wedge \,
  V^{\underline{b}} \right) = - \nabla \left( \overline{\lambda } \, \Gamma_{\underline{ab}} \, \lambda \,
  V^{\underline{a}} \,\wedge \,
  V^{\underline{b}} \right)
\label{rabotaiu}
\end{equation}
The relevance of the above equation in relation with the gauge
symmetries of the quantum action that we are going to consider is
easily explained. As we have already noted, the gauge-fixing condition
we want to implement is the variation with respect to $\delta \Psi$
of the $4$--cycle $\Omega^{[4]}$. Naming $w$ the antighost
which, due to its negative ghost number, can be interpreted as a contraction,
a useful formal way of writing the gauge fermion (\ref{phigaugo})
is the following:
\begin{equation}
  \Phi_{(gauge)} \, = \ft 12 \, i_{\overline{w}} \, \Omega^{[4]} \,
  = \,
   \, \overline{w} \, \Gamma_{\underline{ab}} \, \Psi \, \wedge \,
  V^{\underline{a}} \,\wedge \,
  V^{\underline{b}}
\label{fertile}
\end{equation}
Then equation (\ref{rabotaiu}) guarantees that the transformation:
\begin{equation}
  w \, \mapsto \, w \, + \, a \, \lambda
\label{fantasticone}
\end{equation}
with $a$ an arbitrary parameter of ghost number $-2$ is a symmetry of
the quantum action (\ref{quantaction}), the lagrangian varying by the
total derivative of a term which is also zero on the constrained
surface of pure spinors.
\par
As we are going to see in the following, the BRST invariance of the
quantum action can be achieved if the antighost field $w$ besides (\ref{fantasticone}),
possesses a gauge symmetry which is just a slight modification of the same transformation, namely:
\begin{equation}
  w \, \mapsto \, w \, + \, a \, \mathcal{P}_{q} \, \lambda
\label{fantasticone2}
\end{equation}
where
\begin{eqnarray}
  \mathcal{P}_{(q)}&  \equiv & \frac{1}{2} \left( 1 +{\rm i} {\bf q} { \widehat{\mathbf{\Gamma}}}
\right)\,
\label{kinoteatr}
\end{eqnarray}
is the $\kappa$ supersymmetry projector defined in
eq.(\ref{projector}). This symmetry could be established via the same
chain of arguments we have just pursued if the following two statements were true:
\begin{description}
  \item[A]] The Fierz identity (\ref{basicfierzus}) can be successfully modified to the following one:
\begin{equation}
  \overline{\Psi }\, \mathcal{P}_{(q)} \, \wedge \, \Gamma_{\underline{ab}} \, \Psi \, \wedge \,
  \overline{\Psi} \,\wedge \, \Gamma^{\underline{a}} \, \Psi \, \wedge \,
  V^{\underline{b}} \, = \, 0
\label{basicfierzus2}
\end{equation}
  \item[B]] The $\kappa$--symmetry projector is BRST invariant:
\begin{equation}
  \mathcal{S} \, \mathcal{P}_{(q)}  \, = \, 0
\label{fernandel}
\end{equation}
\end{description}
Indeed, since the operator $\widehat{\mathbf{\Gamma}}$ is actually
proportional to the volume form of the supermembrane, it is
certainly true that $d\mathcal{P}_{(q)}=0$ and condition $\mathbf{B}]$ suffices to
state that:
\begin{equation}
  \mathbf{d}\, \mathcal{P}_{(q)} = 0
\label{dgrassoonP}
\end{equation}
then, in case condition $\mathbf{A}]$ is also true we obtain a new
ghost-form $4$ cycle:
\begin{equation}
  \widehat{\Omega}^{[4]}_{\mathbf{q}} \, = \, \overline{\mathbf{\Psi} }\, \mathcal{P}_{(q)} \, \wedge \, \Gamma_{\underline{ab}} \, \mathbf{\Psi} \, \wedge \,
  \mathbf{V}^{\underline{a}} \,\wedge \,
  \mathbf{V}^{\underline{b}} \quad ; \quad \mathbf{d} \,
  \widehat{\Omega}^{[4]}_{q} \, = \, 0 \quad\mbox{modulo curvatures}
\label{extendedOmegaQ}
\end{equation}
and a new analogous chain of descent equations (\ref{descenteque}):
\begin{eqnarray}
\mathcal{S} \, \Omega^{[p,q-1]}_\mathbf{q} \, + \, d \, \Omega^{[p-1,q]}_\mathbf{q} \, = \,
\Xi^{[p,q]}_\mathbf{q}
\label{descentequeQ}
\end{eqnarray}
leading  in particular to the modified version of
eq.(\ref{rabotaiu}):
\begin{equation}
  \mathcal{S} \, \left(\,2 \,\overline{\lambda }\,\mathcal{P}_{(q)}\, \Gamma_{\underline{ab}} \, \Psi \, \wedge \,
  V^{\underline{a}} \,\wedge \,
  V^{\underline{b}} \right) = - \nabla \left( \overline{\lambda } \, \mathcal{P}_{(q)} \, \Gamma_{\underline{ab}} \, \lambda \,
  V^{\underline{a}} \,\wedge \,
  V^{\underline{b}} \right)
\label{rabotaiu2}
\end{equation}
which guarantees the invariance of the quantum action (\ref{quantaction}) under the
transformation (\ref{fantasticone2}).
\par
It turns out that on the constrained surface of pure spinors and upon
pull-back onto the membrane worldvolume, conditions $\mathbf{A}]$ and $\mathbf{B}]$
are indeed true. To prove it, we rely on the use of  a well-adapted
basis of gamma matrices where we are able to solve the pure spinor
constraints explicitly and henceforth to derive a series of
identities which, at the end of the calculation, we can recast in a fully
$D=11$ covariant form and by this token obtain the desired
verification of the statements we just made. Such a derivation is
discussed in the next section.
\subsection{Pure spinors in a well adapted gamma
 basis and proof of the relevant identities}
\label{welladapted}
The implications of the pure spinor constraints (\ref{costretti_a}-\ref{costretti_b}) are
quite easily discussed and solved if we refer to a gamma matrix basis
which is well adapted to the splitting of the eleven dimensions in
$\mathbf{3}\oplus \mathbf{8}$, the first $\mathbf{3}$ being the dimensions
occupied by the $\mathrm{M2}$-brane worldvolume, the remaining $\mathbf{8}$
being those transverse to the brane. According to this we write
\begin{equation}
  \Gamma_{\underline{a}} \, = \,\left \{\begin{array}{rclcl}
    \Gamma_{{i}} & = & \gamma_i \, \otimes \, T_9 &; & i=0,1,2 \\
    \Gamma_{{2+A}} & = & \mathbf{1} \, \otimes \, T_A &; & A=1,2,\dots,8\
  \end{array} \right.
\label{gammaconstruzia}
\end{equation}
where $\gamma_i$ are $2 \times 2 $ gamma matrices for the $\mathrm{SO(1,2)}$
Clifford algebra, namely:
\begin{equation}
  \left\{ \gamma_i \, , \, \gamma_j\right\} \, = \, 2 \, \eta_{ij} \,
  = \, \mbox{diag}\left\{ +,-,-\right\}
\label{gammapiccole}
\end{equation}
while $T_A$ are $16 \times 16$ gamma matrices for the $\mathrm{SO(8)}$
Clifford algebra with negative metric:
\begin{equation}
\left\{ T_A \, , \, T_B\right\} \, = - \, 2 \, \delta_{AB}
\label{Tamatrici}
\end{equation}
As an explicit representation of the $d=3$ gamma matrices in presence of a mostly
minus metric we can take the following ones in terms of Pauli
matrices:
\begin{equation}
  \gamma_0 \, = \, \sigma_3 \, = \, \left( \begin{array}{cc}
    1 & 0 \\
    0 & -1 \
  \end{array}\right) \quad ; \quad \gamma_1 \, = \, {\rm i} \,
  \sigma_1 \, = \, \left(\begin{array}{cc}
    0 & {\rm i} \\
    {\rm i} & 0 \
  \end{array} \right) \quad ; \quad \gamma_2 \, = \, {\rm i} \,
  \sigma_2 \, = \, \left(\begin{array}{cc}
    0 & 1 \\
    -1 & 0 \
  \end{array} \right)
\label{gammine}
\end{equation}
On the other hand the $\mathrm{SO(8)}$ Clifford algebra with negative metric
admits a representation in terms of completely real and antisymmetric matrices. We
adopt the following one:
\begin{equation}
  T_{A} \, = \,\left \{\begin{array}{rclcl}
    T_{\alpha} & = & \sigma_1 \, \otimes \, \tau_\alpha &; & \alpha=1,2,\dots,7 \\
    T_8 & = & {\rm i} \, \sigma_2 \, \otimes \, \mathbf{1}_{8 \times 8} &; &
    \null
  \end{array} \right.
\label{Tmatricione}
\end{equation}
where $\tau_\alpha$ denotes the $8 \times 8$ completely antisymmetric realization
of the $\mathrm{SO(7)}$ Clifford algebra with negative metric:
\begin{equation}
  \left\{ \tau_\alpha \, , \, \tau_\beta \right\} \, = - \, 2 \, \delta_{\alpha\beta}
  \, \quad ; \quad \tau_\alpha \, = \, - \left( \tau_\alpha \right)^T
\label{taupiccole}
\end{equation}
given by:
{\scriptsize
\begin{eqnarray*}
  \begin{array}{ccccccc}
    \tau_1 & = & \left( \matrix{ 0 & 0 & 0 & 0 & 0 & 0 & 0 & 1 \cr 0 & 0 & 1 & 0 & 0 & 0 & 0 & 0 \cr \
0 & - 1 & 0 & 0 & 0 & 0 & 0 & 0 \cr 0 & 0 & 0 & 0 & 0 & 0 & 1 & 0 \cr 0 & 0 & \
0 & 0 & 0 & - 1 & 0 & 0 \cr 0 & 0 & 0 & 0 & 1 & 0 & 0 & 0 \cr 0 & 0 & 0 & -
                                        1 & 0 & 0 & 0 & 0 \cr -
                                  1 & 0 & 0 & 0 & 0 & 0 & 0 & 0 \cr  }  \right)  & ; & \tau_2 & = & \left(\matrix{ 0 & 0 & -
                                        1 & 0 & 0 & 0 & 0 & 0 \cr 0 & 0 & 0 & \
0 & 0 & 0 & 0 & 1 \cr 1 & 0 & 0 & 0 & 0 & 0 & 0 & 0 \cr 0 & 0 & 0 & 0 & 0 & 1 \
& 0 & 0 \cr 0 & 0 & 0 & 0 & 0 & 0 & 1 & 0 \cr 0 & 0 & 0 & -
                                        1 & 0 & 0 & 0 & 0 \cr 0 & 0 & 0 & 0 & \
- 1 & 0 & 0 & 0 \cr 0 & - 1 & 0 & 0 & 0 & 0 & 0 & 0 \cr  }   \right)  \
\end{array}
\end{eqnarray*}
\begin{eqnarray*}
  \begin{array}{ccccccc}
    \tau_3 & = & \left( \matrix{ 0 & 1 & 0 & 0 & 0 & 0 & 0 & 0 \cr -
                                        1 & 0 & 0 & 0 & 0 & 0 & 0 & 0 \cr 0 & \
0 & 0 & 0 & 0 & 0 & 0 & 1 \cr 0 & 0 & 0 & 0 & -
                                        1 & 0 & 0 & 0 \cr 0 & 0 & 0 & 1 & 0 & \
0 & 0 & 0 \cr 0 & 0 & 0 & 0 & 0 & 0 & 1 & 0 \cr 0 & 0 & 0 & 0 & 0 & -
                                        1 & 0 & 0 \cr 0 & 0 & -
                          1 & 0 & 0 & 0 & 0 & 0 \cr  } \right)  & ; & \tau_4 & = & \left( \matrix{ 0 & 0 & 0 & 0 & 0 & 0 & - 1 & 0 \cr 0 & 0 & 0 & 0 & 0 & -
                                        1 & 0 & 0 \cr 0 & 0 & 0 & 0 & 1 & 0 & \
0 & 0 \cr 0 & 0 & 0 & 0 & 0 & 0 & 0 & 1 \cr 0 & 0 & -
                                        1 & 0 & 0 & 0 & 0 & 0 \cr 0 & 1 & 0 & \
0 & 0 & 0 & 0 & 0 \cr 1 & 0 & 0 & 0 & 0 & 0 & 0 & 0 \cr 0 & 0 & 0 & -
                      1 & 0 & 0 & 0 & 0 \cr  }\right)  \
\end{array}
\end{eqnarray*}
\begin{eqnarray*}
  \begin{array}{ccccccc}
   \tau_5 & = & \left( \matrix{ 0 & 0 & 0 & 0 & 0 & 1 & 0 & 0 \cr 0 & 0 & 0 & 0 & 0 & 0 & -
                                        1 & 0 \cr 0 & 0 & 0 & -
                                        1 & 0 & 0 & 0 & 0 \cr 0 & 0 & 1 & 0 & \
0 & 0 & 0 & 0 \cr 0 & 0 & 0 & 0 & 0 & 0 & 0 & 1 \cr -
                                        1 & 0 & 0 & 0 & 0 & 0 & 0 & 0 \cr 0 & \
1 & 0 & 0 & 0 & 0 & 0 & 0 \cr 0 & 0 & 0 & 0 & - 1 & 0 & 0 & 0 \cr  } \right) & ; & \tau_6 & = & \left( \matrix{ 0 & 0 & 0 & 0 & -
                                        1 & 0 & 0 & 0 \cr 0 & 0 & 0 & 1 & 0 & \
0 & 0 & 0 \cr 0 & 0 & 0 & 0 & 0 & 0 & - 1 & 0 \cr 0 & -
                                        1 & 0 & 0 & 0 & 0 & 0 & 0 \cr 1 & 0 & \
0 & 0 & 0 & 0 & 0 & 0 \cr 0 & 0 & 0 & 0 & 0 & 0 & 0 & 1 \cr 0 & 0 & 1 & 0 & 0 \
& 0 & 0 & 0 \cr 0 & 0 & 0 & 0 & 0 & - 1 & 0 & 0 \cr  } \right)  \
\end{array}
\end{eqnarray*}
\begin{equation}
  \begin{array}{ccc}
    \tau_7 & = & \left( \matrix{ 0 & 0 & 0 & 1 & 0 & 0 & 0 & 0 \cr 0 & 0 & 0 & 0 & 1 & 0 & 0 & 0 \cr \
0 & 0 & 0 & 0 & 0 & 1 & 0 & 0 \cr - 1 & 0 & 0 & 0 & 0 & 0 & 0 & 0 \cr 0 & -
                                        1 & 0 & 0 & 0 & 0 & 0 & 0 \cr 0 & 0 & \
- 1 & 0 & 0 & 0 & 0 & 0 \cr 0 & 0 & 0 & 0 & 0 & 0 & 0 & 1 \cr 0 & 0 & 0 & 0 & \
0 & 0 & - 1 & 0 \cr  } \right)  \
  \end{array}
\label{tauexplicit}
\end{equation}
}
This realization of the $\tau$ matrices admits the following
interpretation:
\begin{equation}
  \left( \tau_{\alpha} \right) _{\beta\gamma} = a_{\alpha\beta\gamma}
   \quad ; \quad \left( \tau_{\alpha} \right) _{\beta 8} = - \left( \tau_{\alpha} \right) _{8
   \beta} \, = \, \delta_{\alpha\beta}
\label{ortofresco}
\end{equation}
where the completely antisymmetric tensor $a_{\alpha\beta\gamma}$
encodes the  structure constants of the octionon algebra or, equivalently corresponds to the components of the unique $\mathrm{G_2}$
invariant $3$--form.
\par
Finally the $16 \times 16$ matrix $T_9$ which anticommutes with all the
$T_A$ has, in this basis, the following structure:
\begin{equation}
  T_9 = - \sigma_3 \, \otimes \, \mathbf{1}_{8\times 8}
\label{T9matra}
\end{equation}
The charge conjugation matrix, with respect to which we have:
\begin{equation}
  C \, \Gamma_{\underline{a}} \, C^{-1} \, = \, - \Gamma_{\underline{a}}^T
\label{chargeconjdef}
\end{equation}
is given by:
\begin{equation}
  C \, = \, \varepsilon \,  \otimes \, \mathbf{1}_{16 \times
  16}\quad ; \quad \left( \varepsilon \, \equiv \, {\rm i} \, \sigma_2
  \,\right)
\label{chargconjval}
\end{equation}
Within this setup we can now address the problem of solving the pure
spinor constraints (\ref{costretti_a}-\ref{costretti_b}). To this end we begin by
parametrizing a \textit{complex} $32$--component spinor $\lambda$ as
the following tensor product:
\begin{equation}
  \lambda = \phi_+ \otimes \zeta_+ \, + \, \phi_- \otimes \zeta_-
\label{tensoreproducto}
\end{equation}
where $T_9 \zeta_\pm = \pm \zeta_\pm$ are $\mathrm{SO(8)}$ spinors of opposite
chiralities and $\phi^\pm$ are $2$-component
$\mathrm{SO(1,2)}$ spinors. Calculating the one-gamma current we obtain:
\begin{equation}
  \overline{\lambda} \, \Gamma_{\underline{a}} \, \lambda \, \equiv
  \, \lambda^T \, C\, \Gamma_{\underline{a}} \, \lambda \, = \, \left
  \{ \begin{array}{lcl}
  \left(  \phi_+^T \, \gamma_i \, \phi_+ \, \right) \zeta_+^T \, \zeta_+ \, - \, \left( \phi_-^T \, \gamma_i \, \phi_-\right)  \, \zeta_-^T \, \zeta_- & ; & i=0,1,2 \\
    2 \, \left( \phi_+^T \, \varepsilon \, \phi_- \right) \, \zeta_+^T \, T_A \zeta_-  & ; & A=1,\dots 8 \
  \end{array} \right.
\label{purecostra}
\end{equation}
Hence the first of the two constraints  (\ref{costretti_a}) can be
easily solved by means of the following two equations:
\begin{eqnarray}
\phi_+ & = & \phi_- = \phi \nonumber\\
\zeta_+^T \, \zeta_+ \, & = &  \zeta_-^T \, \zeta_-
\label{soluzione1}
\end{eqnarray}
which lead to a pure spinor having $23$ independent components. The
argument to count 23 is the following one. By means of
eq.(\ref{soluzione1}) we have explicitly constructed a parametric
solution of the pure spinor constraint equation in terms of an object
which has 32 in general non-zero complex components. We have to see
how many relations are imposed on these 32 components by the
parametric solution. This is easily done. Let us enumerate, according
to the adopted tensor product structure the components of the spinor
(\ref{tensoreproducto}) in the following way:
\begin{equation}
\begin{array}{rclcrcl}
  \lambda^{I} & =& \phi_+^1 \, \zeta_-^I & ;& \lambda^{I+8} & = & \phi_-^1 \, \zeta_+^I   \\
   \lambda^{I+16} & =& \phi_+^2 \, \zeta_-^I & ;& \lambda^{I+24} & = & \phi_-^2 \, \zeta_+^I
 \end{array}
\label{lambdi}
\end{equation}
It is evident that the first of the two equations (\ref{soluzione1})
imposes exactly $8$ equations on the spinor components, namely:
\begin{equation}
  \frac{\lambda^I}{\lambda^{I+16}} \, = \,
  \frac{\lambda^{I+8}}{\lambda^{I+24}} \quad ; \quad
  I=1,\dots 8
\label{ottoeque}
\end{equation}
Hence after the first of (\ref{soluzione1}) has been imposed, we have
$24$ independent components. The second of (\ref{soluzione1}) imposes
just one more condition:
\begin{equation}
  \sum_{I=1}^{16} \, \left[ \left( \lambda^I\right )^2 - \left(
  \lambda^{I+16}\right )^2\right] \, = \, 0
\label{secondacondo}
\end{equation}
which reduces the spinor to $23$ independent
components as it is well known in the literature \cite{Berkovits:2002uc}.
\par
Let us now consider the second constraint  (\ref{costretti_b}). In the following
we will consider the second constraint as a primary constraints and therefore we solve
it in the same spirit as (\ref{costretti_a}).
In the well-adapted gamma matrix basis this reduces to:
\begin{eqnarray}
0 & = & \phi^T \varepsilon \, \gamma_{ij} \, \phi \, \left( \zeta_+^T\, \zeta_+ \, + \, \zeta_-^T\, \zeta_-\right)  \nonumber\\
0 & = & \phi^T \varepsilon \, \gamma_{i} \phi \, \left( \zeta_+^T \,
T_A \, \zeta_- \right)
\label{necostretti}
\end{eqnarray}
which can be easily and uniquely solved by choosing either one of the following four positions:
\begin{equation}
\begin{array}{ccccccccccc}
  \zeta_- & = &\left(\begin{array}{l}
    0\\
    0\
  \end{array} \right)  &; & \zeta_+  & = & \left(\begin{array}{c}
    0 \\
    \omega\
  \end{array} \right) & ; & \omega^T\omega & = &
  0\\
  \null& \null&\null &\mbox{\bf or} &\null & \null & \null & \null &
  \null & \null & \null\\
  \zeta_+ & = &\left(\begin{array}{l}
    0\\
    0\
  \end{array} \right)  &; & \zeta_-  & = & \left(\begin{array}{l}
    \omega \\
    0\
  \end{array} \right) & ; & \omega^T\omega & = &
  0\\
  \null& \null&\null &\mbox{\bf or} &\null & \null & \null & \null &
  \null & \null & \null\\
  \zeta_+ & = &\left(\begin{array}{l}
    0\\
    \omega\
  \end{array} \right)  &; & \zeta_-  & = & \left(\begin{array}{c}
    \omega \\
    0\
  \end{array} \right) & ; & \omega^T\omega & = &
  0\\
  \null& \null&\null &\mbox{\bf or} &\null & \null & \null & \null &
  \null & \null & \null\\
  \zeta_+ & = &\left(\begin{array}{l}
    0\\
    \omega\
  \end{array} \right)  &; & \zeta_-  & = & \left(\begin{array}{c}
    -\omega \\
    0\
  \end{array} \right) & ; & \omega^T\omega & = &
  0\\
  \end{array}
\label{tuttesole}
\end{equation}
where in all four cases  $\omega$ is an
$8$--component complex object with vanishing norm. The manifold of pure
spinors has therefore four disjoint branches given by the four
options in eq.(\ref{tuttesole}). In any case the pure spinor satisfying all the
constraints has $15$ independent components.
\par
In the well adapted basis the world-volume components $\psi_i $ of the
gravitino field which appear in the gauge-fixing
(\ref{diraccus}) can be parametrized as follows:
\begin{equation}
  \psi_i \, = \, \mu_i^+ \, \otimes \, \chi_+ \, + \,  \mu_i^- \, \otimes \,
  \chi_-
\label{lonelypsi}
\end{equation}
where $\mu^\pm_i$ are $3 \times 2$-component vector-spinors of
$\mathrm{SO(1,2)}$ and $\chi^\pm$ are $16$-component spinors of
$\mathrm{SO(8)}$. In this parametrization the gauge-fixing equation
(\ref{diraccus}) reduces to:
\begin{equation}
 \left(  \gamma_{ij} \, \mu_k^+ \, \otimes \, \chi_+ \, - \,  \gamma_{ij} \,
  \mu_k^- \, \otimes \, \chi_- \right) \, \epsilon^{ijk} = 0
\label{2diraccu}
\end{equation}
which implies:
\begin{equation}
  \gamma_i \, \mu^\pm_j \, \eta^{ij} \, = \, 0
\label{muequa}
\end{equation}
Relaying on eq.(\ref{muequa}) we can now prove a Fierz identity which
will be crucial in demonstrating the nilpotency of the BRST operator.
The identity is the following. Define the object:
\begin{equation}
  \Upsilon \, \equiv \, \mathbf{\Gamma}_{\underline{abc}}\, \lambda
  \, \overline{\lambda} \, \mathbf{\Gamma}^{\underline{a}} \, \psi_i
  \, \Pi^{\underline{b}}_j \, \Pi^{\underline{c}}_k \, \epsilon^{ijk}
\label{Ypsilon}
\end{equation}
where $\lambda$ is the pure spinor ghost field and the other items
entering the definition have already been defined above. We want to
show, that independently of the choice of the branch
(\ref{tuttesole}) the structure $\Upsilon$ vanishes modulo the fermionic field
equations, namely upon enforcement of eq.s (\ref{muequa}). To this
effect it suffices to split the sum on the index $\underline{a}$ into
the sum over the first three indices $m$ and over the last eight indices $A$. So we
write:
\begin{eqnarray}
\Upsilon & = & \Upsilon_{[3]} \, + \,  \Upsilon_{[8]} \nonumber\\
\Upsilon_{[3]}  &= & \mathbf{\Gamma}_{pjk}\, \lambda
  \, \overline{\lambda} \, \mathbf{\Gamma}^{p} \, \psi_i
  \, \epsilon^{ijk} \nonumber\\
\Upsilon_{[8]}  & = & \mathbf{\Gamma}_{Ajk}\, \lambda
  \, \overline{\lambda} \, \mathbf{\Gamma}^{A} \, \psi_i
  \,  \epsilon^{ijk} \nonumber\\
\label{orribile}
\end{eqnarray}
and in the well adapted basis we find:
\begin{eqnarray}
  \Upsilon_{[3]} & = & \gamma_{pjk} \, \phi \, \overline{\phi} \, \gamma^p \,
  \mu^+_i \, \epsilon^{ijk} \, \otimes \, T_9 \, \zeta \, \zeta^T \,
  T_9 \, \chi_+ \, + \, \, \gamma_{pjk} \, \phi \, \overline{\phi \,} \gamma^p \,
  \mu^-_i \, \epsilon^{ijk} \, \otimes \, T_9 \, \zeta \, \zeta^T \,
  T_9 \, \chi_- \, \nonumber\\
 & = & - 2 \,{\rm i} \,\left( \phi \, \overline{\phi} \, \gamma^i \,
  \mu^+_i \,   \otimes \, T_9 \, \zeta \, \zeta^T \,
  T_9 \, \chi_+ \, + \, \,  \phi \, \overline{\phi \,} \gamma^i \,
  \mu^-_i \,  \otimes \, T_9 \, \zeta \, \zeta^T \,
  T_9 \, \chi_- \, \right) \nonumber\\
  & \approx & 0 \quad \mbox{modulo field eq.s}
\label{festone}
\end{eqnarray}
which is a consequence of $\gamma_{ijk} \, = \, - \, {\rm i} \, \mathbf{1}_{2
\times 2} \epsilon_{ijk}$.
\par
On the other hand for the second structure we have:
\begin{equation}
  \Upsilon_{[8]} = \gamma_{ij} \, \phi \, \overline{\phi} \, \mu^{+}_k \,
  \epsilon^{ijk} \, \otimes \, T_A \, \zeta \, \zeta^T \, T_A \, \chi_{+} \, +
  \, \gamma_{ij} \, \phi \, \phi \, \mu^{-}_k \,
  \epsilon^{ijk} \, \otimes \, T_A \, \zeta \, \zeta^T \, T_A \chi_{-} \,
\label{pratino}
\end{equation}
Starting from eq (\ref{pratino}) we can show that also
$\Upsilon_{[8]}$ vanishes using the following identity:
\begin{equation}
  T_A \zeta \, \zeta^T \, T_A = \omega^T  \omega \, \,\ft 12 \left( \mathbf{1} - T_9 \right)   \, = \, 0
\label{identitus}
\end{equation}
which is true for all four cases of spinors $\zeta$ listed in
eq.(\ref{tuttesole}).
In this way, independently from the choice of the branch
(\ref{tuttesole}) in the solution of the pure spinor constraints
(\ref{costretti_a}) and (\ref{costretti_b}) we have shown that $\Upsilon =
0$ modulo field equations.
\par
It is interesting to rewrite in a $D=11$ fully covariant way the
result for  off-shell $\Upsilon$. To this end it suffices to
compare the result obtained in equation (\ref{festone}) with the
structure of the $\kappa$--symmetry projector (\ref{projector}) in the
well-adapted basis.  Using
(\ref{gammaconstruzia}) we obtain:
\begin{eqnarray}
  \mathcal{P}_{(q)}&  \equiv & \frac{1}{2} \left( 1 +{\rm i} {\bf q} { \widehat{\mathbf{\Gamma}}}
\right)\, \nonumber\\
& = & \mathbf{1}_{2\times 2} \, \otimes \, \ft 12 \, \left(\mathbf{1}_{16 \times 16} \, + \, T_9 \right)
\label{interpreto}
\end{eqnarray}
On the other hand the second line in eq.(\ref{festone}) can be
rewritten as:
\begin{equation}
  \Upsilon_{[3]} \, = \, - 2 \, {\rm i} \, \mathbf{1}_{2\times 2} \, \otimes \,
  T_9 \, \lambda \, \overline{\lambda} \, \mathbf{\Gamma}^i \, \psi_i
\label{y3new}
\end{equation}
and in view of (\ref{interpreto}) and of the off-shell vanishing of
$\Upsilon_{[8]}$ we can conclude that $\Upsilon$ as defined in eq.(\ref{Ypsilon}) is also equal to the
following expression:
\begin{equation}
  \Upsilon \, = \, 2 \,\widehat{\mathbf{\Gamma}} \, \lambda \,
  \lambda \, \mathbf{\Gamma}^i \, \psi_i
\label{orinale}
\end{equation}
\par
Our next point is to show that in off-shell second order formalism,
namely upon implementation of the algebraic field equation for the
auxiliary fields $\Pi^{\underline{a}}_i$ and $e^i$, without imposing any constraint on the physical fields,
the $\kappa$ supersymmetry projector operator is BRST invariant, namely that eq.(\ref{fernandel}) is true
off-shell. Since the pure spinor ghost is anyhow BRST invariant $\mathcal{S}\lambda =
0$, eq.(\ref{fernandel}) is completely equivalent to the equation
below:
\begin{equation}
  \mathcal{S} \, \left[ \mathcal{P}_{(q )} \, \lambda \right] \, = \, 0 \quad \quad \quad \left( \mbox{off-shell}\right)
\label{Pinvariant}
\end{equation}
which is what we can prove by using the identity (\ref{orinale}). To this
effect let us anticipate a result which we derive in the next section while discussing the BRST invariance of the action.
This latter requires  the  BRST variation of the world-volume
dreibein to be of the following form:
\begin{equation}
  \mathcal{S} \, e^i \, = \, \eta^{im} \, N_{im} \, e^m
\label{Sdreibein}
\end{equation}
with the condition on $N$ to be symmetric $N_{ij} = N_{ji}$.
Using the equations of motion for the auxiliary fields
$\Pi^{\underline{a}}_i$ and $e^i$, namely using the second-order
formalism, we are able to determine $N_{ij}$ explicitly.
Combining eq.(\ref{Vaei}) (\textit{i.e.} the field equation of
$\Pi^{\underline{a}}_i$) with eq.(\ref{SdiVa}) (the
BRST variation of the bulk vielbein), and also eq.(\ref{Sdreibein}), we obtain:
\begin{equation}
 \mathcal{ S} \, \Pi^{\underline{a}}_\ell \, = \, {\rm i} \,
 \overline{\lambda}\, \Gamma^{\underline{a}}\, \psi_\ell\,  - \,
 \Pi^{\underline{a}}_f \, \eta^{fm} \, N_{m\ell}
\label{SofPi}
\end{equation}
Considering next eq.(\ref{flattopullo}), which is just the field equation
of the dreibein in the formulation without $h^{ij}$\footnote{Or can be
alternatively imposed as a gauge fixing condition in the formulation
with  $\mathbb{\mathrm{GL(3,\mathbb{R})}}$ symmetry}, we get:
\begin{equation}
  0 \, = \, \mathcal{S} \, \eta_{ij} \, = \, \mathcal{ S} \, \Pi^{\underline{a}}_i
  \,  \Pi^{\underline{b}}_j \, \eta_{\underline{ab}} \, + \, \mathcal{ S} \,
  \Pi^{\underline{a}}_j
  \,  \Pi^{\underline{b}}_i \, \eta_{\underline{ab}} \,
\label{ornitorinco}
\end{equation}
>From (\ref{ornitorinco}) and (\ref{SofPi}) we immediately obtain:
\begin{equation}
  N_{ij} \, = \, {\rm i} \, \overline{\lambda} \mathbf{\Gamma}_{(i} \, \psi_{j)}
\label{frisco}
\end{equation}
Equipped with these intermediate results we can calculate:
\begin{eqnarray}
\mathcal{S} \, \left[ \mathcal{P}_{(q )} \, \lambda \right] \, & = & \ft 3 2 \, {\mathbf{q}} \, \Upsilon \, + \, \ft 32 \, {\rm i}
\, \mathbf{q} \,
 {\Gamma}_{\underline{abc}} \, \lambda \, \Pi^{\underline{a}}_f \, \, \Pi^{\underline{b}}_f
 \, \, \Pi^{\underline{c}}_r \, \eta^{fm} \, N_{mp} \, \epsilon^{pqr} \nonumber\\
\null & = & \ft 3 2 \, {\mathbf{q}} \, \Upsilon \, + \, 3 \, {\rm i}
\, q \, \widehat{\mathbf{\Gamma}} \, \lambda \, \left( N_{pq} \, \eta^{pq} \, \right
)\nonumber\\
& = &\, - \, 3 \, {\rm i}
\, q \, \widehat{\mathbf{\Gamma}} \, \lambda \, \left( N_{pq} \,
\eta^{pq}\right ) \, + \, 3 \, {\rm i}
\, q \, \widehat{\mathbf{\Gamma}} \, \lambda \, \left( N_{pq} \,
\eta^{pq}\right ) \, = \, 0
\label{fornitore}
\end{eqnarray}
which proves eq. (\ref{Pinvariant}) and hence also
eq.(\ref{fernandel})
The first condition, namely the
Fierz identity (\ref{basicfierzus2}) can be proved in the same well
adapted basis by explicit evaluation for  instance on a computer. It
is just an algebraic identity and it is indeed true.

\subsection{Gauge fixing term and  BRST invariance}
\label{BRSTinvariance}
Relying on the identities shown in the previous sections we can now
prove that the quantum action defined by eq.(\ref{quantaction}), with
the gauge fermion provided by eq.(\ref{fertile}) is indeed BRST
invariant.
\par
According to table \ref{allfieldi} we recall that
the antighost $w$ is just a target space
spinor and we set the following BRST transformation rules:
\begin{equation}
  \mathcal{S} \, w \, = \, \Delta \quad ; \quad  \mathcal{S} \, \Delta \, = \,
  \vartheta
\label{Sonantighost}
\end{equation}
where $\vartheta$ is an object to be determined in such a way that the
final action  be BRST invariant. The gauge fermion  being that in
eq.(\ref{fertile}) the explicit form of the  gauge fixing action is easily evaluated:
\begin{eqnarray}
  \mathcal{A}_{GF} & = & \mathcal{S} \, \int \, \overline{w} \, \Gamma_{\underline{ab}} \, \psi
  \, \wedge \, V^{\underline{a}} \, \wedge \, V^{\underline{b}}
  \nonumber\\
  & = & \int \, \left \{ \, \overline{\Delta} \, \Gamma_{\underline{ab}} \, \psi
  \, \wedge \, V^{\underline{a}} \, \wedge \, V^{\underline{b}} \, -
  \, \overline{w} \, \Gamma_{\underline{ab}} \, \nabla  \lambda
  \, \wedge \, V^{\underline{a}} \, \wedge \, V^{\underline{b}} \,
  \right. \nonumber\\
  & \null & \left. \, + \, 2 \, {\rm i}  \overline{w} \, \Gamma_{\underline{ab}} \,
  \Psi
  \, \wedge \, \overline{\lambda} \, \Gamma^{\underline{a}} \,\Psi \,  \wedge \, V^{\underline{b}}
  \, \right \}
\label{quantumactia}
\end{eqnarray}
If we calculate the BRST transformation of this part of the quantum
action we simply obtain:
\begin{eqnarray}
 \mathcal{S}\, \mathcal{A}_{GF} & = & \int \, \mathcal{S}^2\overline{w} \, \Gamma_{\underline{ab}} \, \Psi
  \, \wedge \, V^{\underline{a}} \, \wedge \, V^{\underline{b}}
=  \int \, \overline{\Psi} \, \Gamma_{\underline{ab}} \, \vartheta
  \, \wedge \, V^{\underline{a}} \, \wedge \,
  V^{\underline{b}}\nonumber\\
  & = & \int \, \overline{\Psi} \, \mathbf{\Gamma}_{ij} \, \vartheta
  \, \wedge \, e^{i} \, \wedge \, e^{j} \nonumber\\
  &=& - \ft 16 \, \int \, \overline{\psi}_k \, \mathbf{\Gamma}_{ij} \,\vartheta \,
  \epsilon^{kij} \, \mbox{Vol}(3)
\label{SAGF}
\end{eqnarray}
where  we have used the notation $ \mbox{Vol}(3)
=\epsilon_{\ell_1\ell_2\ell_3} \, e^{\ell_1} \, \wedge \, e^{\ell_2}
\, \wedge \, e^{\ell_3} $ already introduced before.
The question is whether there exists a $\vartheta$ appropriate to
cancel the BRST variation of the classical action. The idea behind
this procedure is that the nilpotency of the BRST operator is
preserved if eq.(\ref{nilpotentcirca}) is true on all fields. Hence,
in view of our previous discussions, $\vartheta$ should be one of the
gauge symmetries of the antighost $w$, namely, either
$\vartheta$ should be proportional to $\lambda$ or to $\mathcal{P}_{q}\,
\lambda$. We will explicitly demonstrate that the second is the right
choice. In both cases the results obtained in the previous sections
already guarantee that:
\begin{equation}
  \mathcal{S}^3 w = \mathcal{S} \, \vartheta \, = \, 0
\label{cuffo}
\end{equation}
as it should be for consistency.
\par
So let us now calculate the BRST variation of the classical action.
 Here we use the $1.5$ order formalism, namely we vary
the first order action (\ref{classaction}), but we consider only the
variation of the physical fields, ignoring that of the auxiliary
fields. Then after variation we implement, for the auxiliary fields
their value set by their own field equation.
So, for the classical
action we find:
\begin{eqnarray}
  \mathcal{S} \, \mathcal{A}_{class}& = & \int \, \Big\{ {\rm i} \,
  \Pi^i_{\underline{a}} \overline{\Psi} \, \Gamma^{\underline{{a}}} \, \lambda \,
  \wedge \, e^j \, \wedge \, e^k \, \epsilon_{ijk} \, - \, \mathbf{q}
  \, \overline{\Psi} \, \Gamma_{\underline{ab}} \, \lambda \,
  \wedge \, V^{\underline{a}} \, \wedge \, V^{\underline{b}} \,
  \Big\}
  \nonumber\\
  & = & \int \, \Big \{ {\rm i} \, \overline{\Psi} \,
  \mathbf{\Gamma}^i \, \lambda \, \wedge \, e^j \, \wedge \, e^k \,
  \epsilon_{ijk} \, - \, \mathbf{q} \, \overline{\Psi} \, \mathbf{\Gamma}_{ij}
  \, \lambda \, \wedge \, e^i \, \wedge \, e^j \, \Big\}
  \nonumber\\
  & = & \int \, {\rm i} \ft 23 \, \overline{\lambda} \,
  \mathbf{\Gamma}_i \, \mathcal{P}_q \, \psi_j \, \eta^{ij} \,
  \mbox{Vol}(3)
\label{varioclassa}
\end{eqnarray}
where  we have used the identity:
\begin{eqnarray}
 {\rm i} \, \ft 23 \,\eta^{ij} \, \overline{\lambda}  \mathbf{\Gamma}_i \, \mathcal{P}_q \, \psi_j & = & {\rm i}\, \ft 13 \,
\overline{\lambda}  \mathbf{\Gamma}_i \, \psi_j  \, - \, \mathbf{q} \, \ft 16 \, \overline{\lambda} \, \mathbf{\Gamma}_{ij} \, \psi_m \, \epsilon^{ijm} \nonumber\\
\null & = & - {\rm i} \, \ft 23 \, \eta^{ij} \, \overline{\psi}_i  \mathbf{\Gamma}_j \, \mathcal{P}_q \,
\lambda
\label{identitu}
\end{eqnarray}
which follows from standard gamma matrix manipulations. Similarly one
can prove the other identity:
\begin{equation}
 \ft 13 \,  \overline{\psi}_k \, \mathbf{\Gamma}_{ij} \,\mathcal{P}_q
 \, \lambda
 \, =\, {\rm i} \mathbf{q} \,\eta^{ij} \, \overline{\psi}_i  \mathbf{\Gamma}_j \, \mathcal{P}_q \,
\lambda
\label{identitu2}
\end{equation}
Combining these results we conclude that it suffices to set:
\begin{equation}
  \vartheta = 4 \, \mathcal{P}_q \, \lambda
\label{office}
\end{equation}
which is consistent with our previous statements. Indeed it
corresponds to a shift symmetry of the antighost and satisfies the
closure condition (\ref{cuffo})

\subsection{Primary, Secondary and Pure Spinor Constraints}

Before closing this section, we have  to spend some words concerning
the constraint structure of the theory and for that we use the Dirac procedure.

In contrast to \cite{Berkovits:2002uc}, we adopt two types of constraints
from the beginning: (\ref{costretti_a}) and (\ref{costretti_b}) since
they are implied by the supersymmetry algebra. So, we consider them as
primary constraints. Those two constraints are first class
constraints since they commute (using the Poisson brackets)
\begin{equation}\label{fccA}
\{\bar\lambda \Gamma^{\underline a} \lambda,  \bar\lambda \Gamma^{\underline b} \lambda\} =0\,, ~~~~
\{\bar\lambda \Gamma^{\underline a} \lambda,  \bar\lambda \Gamma^{\underline bc} \lambda \Pi_{\underline c i}\} =0\,, ~~~~
\{\bar\lambda \Gamma^{\underline ad} \Pi_{\underline d j} \lambda,
\bar\lambda \Gamma^{\underline bc} \lambda \Pi_{\underline c i}\} =0\,,
\end{equation}
where we have assumed that $\{\Pi^{\underline a}_{i}, \Pi^{\underline b}_{j}\}=0$
(since $ \Pi^{\underline b}_{i}$ is the conjugate momentum to $x^{\underline a}$).  In addition, using the
Fierz identities one can prove that
they are BRST invariant. Nevertheless,
the action ${\cal A}_{quantum}$ is non-linear and therefore the primary
constraints yield secondary constraints by computing the commutator between
the Hamiltonian and the primary constraints. However,
since we want to stick to Lagrangian formalism, we can check whether the action
is invariant under the gauge symmetries generated by the primary constraints.
We found that there are secondary constraints which are differential
constraints and automatically implemented by imposing the field
equations. Furthermore, as we have already seen in previous sections,
the only gauge symmetries which do not impose any further constraints
are those which are needed to close the BRST algebra (\ref{cuffo}).

We have to enumerate some differences with the previous approach
using the Hamiltonian formalism \cite{Berkovits:2002uc}.
We have seen that the naive covariantization of the field equations for $\theta^{\underline \alpha}$
\begin{equation}\label{fccB}
\partial_{0} \theta^{\underline \alpha} + \epsilon^{IJ} \Gamma^{\underline a\alpha}_{\underline \beta} \Pi_{\underline a I}\partial_{J} \theta^{\underline \beta} =0 \quad
\longrightarrow \quad
\gamma^{ij}\partial_{j} \theta^{\underline \alpha} +
\epsilon^{ijk} \Gamma^{\underline a\alpha}_{\underline \beta}
\Pi_{\underline a j}\partial_{k} \theta^{\underline \beta} =0
\end{equation}
is not consistent.
The reason
is that the new equation imposes too strong constraints on the fields
$\theta$'s which has only constant solution. Therefore, we decided
to use the Dirac equation for the second half of the $\theta$'s (those
which would have been projected away by the $\kappa$-symmetry) and
we get a more complicate action. The way to see that the number of
conjugate fields $w_{\underline\alpha}$ is correct is to check that
the wave operator is a quadratic matrix in spinor representation.

It is interesting to compare the supermembrane action with
the superstring action quantized with the pure spinors
\cite{Berkovits:2000fe}. In particular, we compare the
``gauge fixing part'' ${\cal S} \int \Phi_{gauge}$ in Hamiltonian and
Lagrangian formalism (in the Hamiltonian formalism the BRST differential
operator ${\cal S}$ is replaced by the BRST charge $Q$).
We recall that the gauge fixing term for the superstring (we
consider here type IIA to compare with the supermembrane
action) in the worldsheet light-cone coordinates reads as follows
\begin{equation}\label{fccC}
{\cal A}_{gauge} =Q \int d^{2}z \left(w_{\alpha L z} \bar\partial \theta^{\alpha}_L +  w^{\alpha}_{R \bar z}
\partial \theta_{\alpha, R}\right)\,.
\end{equation}
This part action admits two different interpretations. On one hand,
we can see (\ref{fccC}) as a gauge-fixed version of the following
action
\begin{equation}\label{fccD}
{\cal A}_{gauge} ={\cal S} \int d\sigma d\tau \Big(
w_{L}^{i} ( \eta_{ij} + \epsilon_{ij}) d\theta_{L} \wedge e^j +
w_{R}^{i} ( \eta_{ij} - \epsilon_{ij}) d\theta_{R} \wedge e^j
\Big)\,.
\end{equation}
where the second components for $w_{\alpha L z}$ and for
$w^{\alpha}_{R \bar z}$ have been introduced.
The two projectors $ (\eta_{ij} \pm \epsilon_{ij})$ imply that
the action (\ref{fccD}) is invariant under $\delta w_{L/R}^{i} = (\eta_{ij} \mp \epsilon_{ij}) \varphi_j$.
On the other hand, we obtain the action (\ref{fccC})
by dimensional reduction from the supermembrane action
\begin{equation}\label{fccE}
{\cal A}_{gauge} ={\cal S} \int d^{3}x \Big(
w \, \epsilon^{ijk} \Pi^{\underline b}_{i} \Pi^{\underline a}_j \Gamma_{\underline ab}
\partial_k \theta
\Big)\,.
\end{equation}
First, we split the 11d indices $\underline a$ into 10d indices $(a, 11)$ and then,
we integrate
over the third worldvolume coordinate. This eliminates the contribution
coming from $\Gamma_{ab}$ and we use the usual gamma matrix identifications
$\Gamma_{11} = \gamma_3 \otimes T_9$ and
$\gamma_i = \gamma_3 \epsilon_{ij} \gamma^j$. Thus,
we are left with the 2-forms terms
\begin{equation}\label{fccE_bis}
{\cal A}_{gauge} ={\cal S} \int d\sigma d\tau
\Big(
w_L \gamma_a d\theta_L - w_R \gamma_a d\theta_R \Big)\wedge V^a\,,
 \end{equation}
where we have redefined the antighost fields $w \rightarrow (w_{L}, w_{R})$
in a suitable way. It is obvious to compare this action with the string action
with the pure spinors. In this way, we see that the antighosts of superstrings
can also be represented by a spinor rather than by a vector spinor $w_{L/R}^{i}$
as it is mandatory in the supermembrane. The formulation in term of a single
spinor antighost is natural from the viewpoint of FDA. Furthermore, one can
also see that the pure spinor constraints (\ref{costretti_a})-(\ref{costretti_b})
used in the present framework are directly related to the pure spinor constraints
of superstring type IIA. We plan to discuss this point further in a future work.


\section{Examples: the quantum action of the supermembrane on specific backgrounds}
\label{exempla}
The construction that we have described in the previous sections
provides explicit formulae for the supermembrane quantum action in terms of target space
supercoordinates any time we have at our disposal an explicit
parametrization of the $D=11$ superspace geometry. All what we need
are just three geometrical data:
\begin{enumerate}
  \item The supervielbein one--form $V^{\underline{a}}$
  \item The gravitino one-form $\Psi$
  \item The three--form $A^{[3]}$
\end{enumerate}
As an illustration of the ultimate content of our result we consider
two explicit cases, namely flat D=11 superspace and the $\mathrm{AdS_4} \times
\mathrm{S}^7$ solution. 

We remind the reader that the pure spinor superstring on arbitrary background has been 
studied in \cite{Berkovits:2001ue} where it is shown that the pure spinor conditions and the holomorphicity conditions 
imply the supergravity constraints for heterotic and type IIA/B superstrings. The background 
formulation of pure spinor supermembrane is analyzed in \cite{Berkovits:2002uc} in Hamiltonian 
formalism. On the other hand, the arbitrary background formulation of $\kappa$-symmetric $p$-branes 
has been initiated in \cite{Dall'Agata:1999wz,Billo:1999ip} for $M2$ and the superparticle 
and then extended to any $p$-brane including also worldvolume gauge fields in 
\cite{Fre:2002kc,Fre:2002jf}.

\subsection{$D=11$ flat superspace}
In this case the representation of the needed supergeometrical data
is very simple. Naming $X^{\underline{a}}$ the eleven bosonic
coordinates and $\theta$ the $32$--component Majorana spinor, the
structural equations (\ref{FDAcompleta}) with zero curvatures are
immediately solved by setting:
\begin{eqnarray}
V^{\underline{a}} & = & dX^{\underline{a}} \, + \, {\rm i} \, \overline{\theta} \, \Gamma^{\underline{a}}
d\theta \nonumber\\
\Psi & = & d\theta \nonumber\\
\mathbf{A}^{[3]} & = & \ft 12 \, \overline{\theta} \,
\Gamma_{\underline{ab}}\, d\theta \, \wedge \, dX^{\underline{a}} \, \wedge \,
dX^{\underline{b}} \, + \, {\rm i} \, \ft 14 \,\overline{\theta} \,
\Gamma_{\underline{ab}}\, d\theta \, \wedge \, \overline{\theta} \, \Gamma^{\underline{a}}
d\theta \, \wedge \, dX^{\underline{b}} \nonumber\\
& \null & - \ft {1}{12} \, \overline{\theta} \,
\Gamma_{\underline{ab}}\, d\theta \, \wedge \,\overline{\theta} \, \Gamma^{\underline{a}}
d\theta \, \wedge \,\overline{\theta} \, \Gamma^{\underline{b}}
d\theta
\label{A3piatto}
\end{eqnarray}
In proving that the last line of eq.(\ref{A3piatto}) does indeed
satisfy the required relation:
\begin{equation}
  d\mathbf{A}^{[3]}= \ft 12 \, \overline{\Psi} \, \wedge
  \Gamma_{\underline{ab}} \Psi \, \wedge \, V^{\underline{a}} \, \wedge \, V^{\underline{b}}
\label{opla}
\end{equation}
one has just to rely on the following Fierz identity:
\begin{equation}
  \overline{\theta} \,
\Gamma_{\underline{ab}}\, d\theta \, \wedge \, d\overline{\theta} \, \wedge \, \Gamma^{\underline{a}}
d\theta = - d \overline{\theta} \, \wedge \,
\Gamma_{\underline{ab}}\, d\theta \, \wedge \, \overline{\theta} \, \Gamma^{\underline{a}}
d\theta
\label{nuovaide}
\end{equation}
Substituting the above formulae into the supermembrane action it becomes
fully explicit in terms of all the fields.
\subsection{The supermembrane on $\mathrm{AdS_4} \times \mathbb{S}^7$}
Another interesting case of backgrounds on which the quantum action
of the supermembrane can be considered is provided by the
Freund-Rubin solutions of $D=11$ supergravity of type\footnote{For a review of Kaluza-Klein compactifications of M--theory
we refer the reader to \cite{castdauriafre}}
\begin{equation}
  \mathcal{M}_{11}\, = \, \mathrm{AdS}_4 \, \times \, \mathrm{\frac GH}
\label{M11GH}
\end{equation}
where $\mathrm{AdS}_4$ denotes $4$--dimensional anti de Sitter space
and $\mathrm{G/H}$ is a $7$--dimensional coset manifold equipped with an invariant Einstein metric
and admitting $\mathcal{N}$ Killing spinors $\eta_A$. Naming $y$ the
coordinates of such a $7$-manifold its vielbein and spin-connection one--forms are
respectively denoted $\mathcal{B}^\alpha(y),
\mathcal{B}^{\alpha\beta}(y)$ and, in order to solve the D=11 field equations, they satisfy the
following  structural equations:
\begin{eqnarray}
d \mathcal{B}^\alpha \, + \,\mathcal{B}^{\alpha\beta} \, \wedge \, \mathcal{B}^\gamma \, \eta_{\beta\gamma}
 & = & 0 \nonumber\\
d\mathcal{B}^{\alpha\beta} \, + \, \,\mathcal{B}^{\alpha\gamma} \, \wedge \, \mathcal{B}^{\delta\beta} \,
\eta_{\beta\gamma}  & = &
\mathcal{R}^{\alpha\beta}_{\phantom{\alpha\beta}\gamma\delta} \,
\mathcal{B}^\gamma \, \wedge \, \mathcal{B}^\delta \nonumber\\
\mathcal{R}^{\alpha\beta}_{\phantom{\alpha\beta}\gamma\delta} & = &
\, 12 \, e^2
\delta^{\alpha}_\beta
\label{internal}
\end{eqnarray}
where $e$ is named the Freund-Rubin parameter and it is the only scale
parameter of the entire solution. In terms of these objects the
Killing spinors are eight component spinors of the tangent group
$\mathrm{SO(7)}$ that are required to satisfy the following equation:
\begin{equation}
  d \eta_a \, + \, \ft 14 \, \mathcal{B}^{\alpha\beta} \,
  \tau_{\alpha\beta} \, \eta_A = e \, \mathcal{B}^{\alpha} \,
  \tau_{\alpha} \, \eta_A \, \quad ; \quad (A=1,\dots, \mathcal{N})
\label{Kspi}
\end{equation}
In eq.(\ref{Kspi}), by $\tau_\alpha$  we have denoted the seven dimensional $8 \times 8$
gamma matrices already introduced in section \ref{welladapted} which
satisfy the standard Clifford algebra with negative metric
$\eta_{\alpha\beta} = -\delta_{\alpha\beta}$. The complete symmetry
of the solution (\ref{M11GH}) is given by the supergroup:
\begin{equation}
 \mathcal{SG} = \mathrm{Osp(\mathcal{N} | 4)} \, \times \, \mathrm{N}_\mathrm{G}\left (\mathrm{SO(\mathcal{N})} \right )
\label{supergruppo}
\end{equation}
where $\mathrm{N}_\mathrm{G}\left (\mathrm{SO(\mathcal{N})} \right )$
denotes the normalizer of the $\mathrm{R}$-symmetry group
$\mathrm{SO(\mathcal{N})}$ within the full isometry group $\mathrm{G}$ of the
internal seven manifold. In the book \cite{castdauriafre} it was shown how to construct the
full $D=11$ superspace geometry corresponding to this class of solutions in terms of
the geometrical data specified above and in terms of the super
geometry of the super coset manifold:
\begin{equation}
  \mathcal{SM} \, = \, \frac {\mathrm{Osp(\mathcal{N} | 4)}}{ \mathrm{SO(1,3)}
  \times \mathrm{SO(\mathcal{N})} }
\label{feluche}
\end{equation}
Consider the Maurer-Cartan equations of the supergroup $\mathrm{Osp(\mathcal{N} |
4)}$ which can be written as follows:
\begin{eqnarray}
d \omega^{ab} + \omega^{ac} \, \wedge \,\omega_c{}^b + 16 e^2 E^a \,
\wedge \,E^b &=&
-{\rm i}\,  2 e \,  \overline{\Psi}_A \, \wedge \gamma^{ab} \gamma^5 \psi_A, \nonumber \\
d E^a + \omega^a_{\phantom{a}c}\, \wedge \, E^c &=& {\rm i} \ft 12 \, \overline{\psi}_A \, \wedge \,\gamma^a
\psi_A, \\
d \psi_A - \frac{1}{4} \omega^{ab}\, \wedge \, \gamma_{ab} \psi_A
+ e {\mathcal{A}}_{AB} \, \wedge \,\psi_B &=&  2 e \,
 E^a \, \wedge \,\gamma_a \gamma_5 \psi_A, \nonumber\\
d {\mathcal{A}}_{AB} + e  {\mathcal{A}}_{AC}\, \wedge \, {\mathcal{A}}_{CB} &=& 4 \, {\rm i}  \overline{\psi}_A \,
\wedge \,\gamma_5 \psi_B, \nonumber\\
\label{orfan25}
\end{eqnarray}
where $\omega^{ab}$, the spin--connection, is dual to the generators
$J_{ab}$ of $\mathrm{SO(1,3)}$, $E^a$, the vielbein is dual to the translation
generators $P_a$  of $\mathrm{SO(2,3)}$, ${\mathcal{A}}_{AB}$, the $R$-symmetry
connection is dual to the generators $T_{AB}$ of $\mathrm{SO(\mathcal{N})}$
and finally $\psi_A$, the gravitino, is dual to the $\mathcal{N}$
supersymmetry charges $Q_A$. Suppose that you have constructed a
solution of these Maurer-Cartan equations on the coset (\ref{feluche})
using some parametrization of the latter in terms of four bosonic
coordinates $x^a$ and $4 \times \mathcal{N}$ fermionic coordinates
$\theta_A$. Then the general recipe to construct the $D=11$ vielbein
$V^{\underline{a}}$ and the $D=11$ gravitino $\Psi$ is the following
\cite{castdauriafre}
\begin{equation}
  \begin{array}{rcl}
    V^{\underline{a}} & = & \left \{ \begin{array}{rcl}
      V^a & = & E^a \\
      V^\alpha & = & \mathcal{B}^\alpha \, + \, \ft 18 \sum_{A,B} \,\overline{\eta}_A \, \tau^\alpha \, \eta_B \,
      {\mathcal{A}}_{AB}\\
    \end{array} \right.\\
    \Psi & = & \sum_{A} \, \eta_A \, \otimes \, \psi_A
  \end{array}
\label{supervielbein}
\end{equation}
The only other item which is necessary in order to write the explicit
form of the supermembrane  action  is an explicit
parametrization of the three-form $A^{[3]}$. As it was noted in
\cite{castdauriafre}, while the curvature $\mathbf{F}^{[4]}$ can be written
intrinsically in terms of the above geometrical data for any
background of the considered type, the corresponding gauge potential
$\mathbf{A}^{[3]}$ can be explicitly solved only within an explicit
parametrization of the supercoset (\ref{feluche}). In \cite{Dall'Agata:1999wz}
the solution of this problem was explicitly found for the case of the
seven sphere by using the so named supersolvable parametrization of the coset. We
refer the reader to \cite{Dall'Agata:1999wz} for all the details and
we just quote the result.
The bosonic coordinates of $\mathrm{AdS_4}$ are named ($\rho$, $t$, $w$, $x$)
and the $\mathrm{AdS}$ metric is written as follows:
\begin{equation}
\label{nearhor}
ds^2 = \rho^2 \left( -dt^2 + dx^2 + dw^2 \right)
+  \frac{R^2}{4} \frac{1}{\rho^2} d\rho^2
\end{equation}
The parametrizations of the vielbeins in
terms of these bosonic coordinates and
of the eight four--dimensional fermionic ones
($\theta_{\alpha}^A$) is the following
\begin{eqnarray}
E^0 &=& - \rho dt - 2e \rho \overline{\theta}^A \gamma^0 d \theta^A, \nonumber \\
E^1 &=& \rho dw - 2e \rho \overline{\theta}^A \gamma^1 d \theta^A, \nonumber \\
E^2 &=& \rho dx - 2\, e \, \rho \overline{\theta}^A \gamma^3 d \theta^A,
\nonumber\\
E^3 &=& \frac{R}{2} \frac{1}{\rho} d\rho,  \label{param1}
\end{eqnarray}
and for the gravitino we have:
\begin{equation}
\psi^A = \sqrt{2e\, \rho}
\; \left( \begin{tabular}{c} $0$ \\ $0$ \\ $d\theta_1^A$ \\
$d \theta_2^A$ \end{tabular} \right),
\end{equation}
where $\displaystyle \theta^A = \frac{1-\gamma^5\gamma^2}{2} \theta^A$ and
$\overline{\theta}^A = \theta^A \gamma^0$.
It can also be found that the $\mathrm{SO(8)}$ connection $\mathcal{A}$, in this
parametrization, is identically zero:
\begin{equation}
{\cal A}_{AB} = 0.
\label{orfan11}
\end{equation}
and in this parametrization the three-form can be written as follows:
\begin{equation}
A^{[3]} = \ft 16 \, E^i\, \wedge \, E^j\, \wedge \, E^k
\,\epsilon_{kji}\, - \, \frac{1}{2e} \, \sum_{A,B} \,
\mathcal{B}^{\alpha} \, \wedge \, \eta_A \tau_{\alpha} \eta_B \;\overline{\psi}_A \, \wedge
\,\psi_B.
\label{orfan29}
\end{equation}
where the index $i=0,1,2$.
\par
Using these data inside the general formulae presented in this
article the quantum action of the supermembrane on the background $\mathrm{AdS_4 }\times
\mathrm{S}^7$, becomes fully explicit.


\section{Conclusions}
\label{concludo}
We have shown that the pure spinor supermembrane has a nice geometrical structure and
we have shown how to use it in order to consider the theory on any background.
The geometrical structure uncovered by our analysis is deeply rooted
in the structure of the Free Differential Algebra of M-theory and
eventually in the cohomology structure of the $D=11$ super Poincar\'e
algebra from which the FDA streams.
\par
However, this is only a starting point towards a more complete analysis. There are several
problems that can be tackled using the action presented here. Just to mention some of them:
\begin{enumerate}
  \item One loop computations
of the supermembrane instanton contributions to the 11d superpotential (see \cite{Harvey:1999as}),
  \item Compactifications on manifolds of $\mathrm{G_2}$ holonomy or of weak $\mathrm{G_2}$-holonomy,
  \item Amplitudes and spectrum.
  \item Extension of our methods to other $p$-branes, in particular
  the $D3$-brane and the $M5$-brane
\end{enumerate}
 We hope to report on some of them soon.

\appendix
\section{ A Note on Free Differential Algebras}
\label{appA}
The algebraic structure that goes
under the name of Free Differential Algebra
 was independently discovered at the beginning of the eighties in Mathematics by
Sullivan \cite{sullivan} and in Physics by one of the present authors (P.F.) in collaboration with R. D'Auria
\cite{Fdorinal}. Free Differential Algebras (FDA) are a  categorical extension of the
notion of Lie algebra and constitute the natural mathematical
environment for the description of the algebraic structure of higher
dimensional supergravity theory, hence also of string theory. The
reason is the ubiquitous presence in the spectrum of
string/supergravity theory of antisymmetric gauge fields ($p$--forms)
of rank greater than one. The very existence of FDA.s is a
consequence of Chevalley cohomology of ordinary Lie algebras and
Sullivan has provided us with a very elegant classification scheme of
these algebras based on two structural theorems rooted in the set up
of such an elliptic complex. As it was already noted  about two decades ago
in \cite{comments},  FDA.s have the additional fascinating property that,
differently from ordinary Lie algebras they already encompass their
own gauging. Indeed the first of Sullivan's structural theorems, which is
in some sense analogous to Levi's theorem for Lie algebras, states
that the most general FDA is a semi-direct sum of a so called minimal
algebra $\mathbb{M}$ with a contractible one $\mathbb{C}$. The
generators of the minimal algebra are physically interpreted as the
connections or \textit{potentials}, while the contractible generators
are physically interpreted as the \textit{curvatures}. The real
hard--core of the FDA is the minimal algebra and it is obtained by
setting the contractible generators (the curvatures) to zero. The
structure of the minimal algebra $\mathbb{M}$, on its turn, is beautifully
determined by Chevalley cohomology of $\mathbb{G}$. This happens to
be the content of Sullivan's second structural theorem. A recent review of FDAs also in relation
with compactifications is contained in \cite{chevrolet}. Other recent work on the topic
is contained in \cite{Bandos:2005mm} and in 
\cite{Hatsuda:2000mn}.


\end{document}